\documentclass[a4paper,USenglish]{lipics-v2018}

\usepackage{xpatch}
\makeatletter
\xpatchcmd{\@evenhead}{\textcolor{blue}{XX}\else\@ArticleNo\fi:\thepage}{\else\@ArticleNo:\fi\thepage}{}{}
\xpatchcmd{\@oddhead}{\textcolor{blue}{XX}\else\@ArticleNo\fi:\thepage}{\else\@ArticleNo:\fi\thepage}{}{}
\makeatother

\title{
\mbox{A population protocol for exact majority with}
\mbox{$O(\log^{5/3} n)$ stabilization time and asymptotically}
\mbox{optimal number of states}
}

\titlerunning{Population protocols for exact majority}

\author{Petra Berenbrink}{Universität Hamburg, Germany}{petra.berenbrink@uni-hamburg.de}{}{}
\author{Robert Elsässer}{University of Salzburg, Austria}{elsa@cosy.sbg.ac.at}{}{}
\author{Tom Friedetzky}{Durham University, U.K.}{tom.friedetzky@dur.ac.uk}{}{}
\author{Dominik Kaaser}{Universität Hamburg, Germany}{dominik.kaaser@uni-hamburg.de}{}{}
\author{Peter Kling}{Universität Hamburg, Germany}{peter.kling@uni-hamburg.de}{https://orcid.org/0000-0003-0000-8689}{}
\author{Tomasz Radzik}{King's College London, U.K.}{tomasz.radzik@kcl.ac.uk}{}{Tomasz Radzik's work has been supported by EPSRC grant EP/M005038/1, ``Randomized algorithms for computer networks''.}

\authorrunning{P. Berenbrink, R. Elsässer, T. Friedetzky, D. Kaaser, P. Kling, and T. Radzik}

\Copyright{Petra Berenbrink, Robert Elsässer, Tom Friedetzky, Dominik Kaaser, Peter Kling, Tomasz Radzik}

\subjclass{
\ccsdesc[500]{Theory of computation~Distributed computing models}
}

\keywords{Population Protocols, Randomized Algorithms, Majority}

\category{}

\relatedversion{}

\nolinenumbers 
\hideLIPIcs  

\let\originalparagraph\paragraph
\def\paragraph#1{\originalparagraph*{#1.}}
\usepackage{microtype}

\usepackage{graphicx}
\usepackage[misc,geometry]{ifsym}
\usepackage{verbatim}
\usepackage{amsfonts}
\usepackage{caption}
\usepackage{tabularx}
\usepackage{array}
\usepackage{booktabs}
\usepackage{multirow}
\usepackage{fancyvrb}
\usepackage{amsmath}
\usepackage{algorithm2e}
\usepackage{hyperref}
\usepackage{framed}
\usepackage{xspace}

\usepackage{enumerate}

\usepackage{latexsym}
\usepackage{amsmath,amssymb}

\usepackage{float}

\def\a{\alpha} \def\b{\beta} \def\d{\delta} \def\D{\Delta}
    \def\g{\gamma}
\def\G{\Gamma}  
\def\ep{\epsilon}

\def\t{\tau}

\def\cT{{\cal T}}

\def\cE{{\cal E}}

\newcommand{\Let}{\leftarrow}
\newcommand{\fail}{\mbox{\it fail}}
\newcommand{\token}{\mbox{\it token}}
\newcommand{\notoken}{\emptyset}
\newcommand{\true}{\mbox{\it true}}
\newcommand{\false}{\mbox{\it false}}

\newcommand{\doubled}{\mbox{\it doubled}}
\newcommand{\vtoken}{\mbox{\it token}}
\newcommand{\Time}{\mbox{\it time}}

\newcommand{\done}{\mbox{\it done}}
\newcommand{\normal}{\mbox{\it normal}}
\newcommand{\outofsync}{\mbox{\it out\_of\_sync}}

\newcommand{\phase}{\mbox{\it phase}}
\newcommand{\epoch}{\mbox{\it epoch}}
\newcommand{\EpochInv}{\mbox{\it EpochInvariant}}

\newcommand{\PhaseInvA}{\mbox{\it PhaseInvariant1}}

\newcommand{\additionalepoch}{\mbox{\it additional\_epoch}}
\newcommand{\epochpart}{\mbox{\it epoch\_part}}
\newcommand{\ageinepoch}{\mbox{\it age\_in\_epoch}}

\newcommand{\phasestep}{\mbox{\it phase\_step}}
\newcommand{\epochstep}{\mbox{\it epoch\_step}}
\newcommand{\stage}{\mbox{\it stage}}
\newcommand{\doubling }{\mbox{\it doubling}}
\newcommand{\canceling }{\mbox{\it canceling}}
\newcommand{\beginningphase}{\mbox{\it beginning}}
\newcommand{\middlephase}{\mbox{\it middle}}
\newcommand{\phaseend}{\mbox{\it ending}}

\newcommand{\consistent}{\mbox{\it Consistent}}
\newcommand{\nextstep}{\mbox{\it NextStep}}
\newcommand{\polylog}{\mbox{\rm polylog}\,}
\newcommand{\ExactMajority}{\mbox{\tt Majority}\xspace}
\newcommand{\FastExactMajorityA}{\mbox{\tt FastMajority1}\xspace}
\newcommand{\FastExactMajorityB}{\mbox{\tt FastMajority2}\xspace}

\def\E{\mathbf{E}}

\def\Pr{\mbox{{\bf Pr}}}
\def\whp{{\it w.h.p.}\xspace}
\def\wlp{{\it w.l.p.}\xspace}
\def\Whp{{\it W.h.p.}\xspace}
\def\wrt{{w.r.t.}\xspace}
\def\Wrt{{W.r.t.}\xspace}

\def\cF{{\cal F}}

{\unskip\nobreak\hskip 1em plus 1fil\nobreak$\Box$
\parfillskip=0pt%
\endtrivlist}
{\unskip\nobreak\hskip 2em plus 1fil\nobreak$\Box$
\parfillskip=0pt%
\endtrivlist}
\newenvironment{proofof}[1]{\trivlist\item[]{{\bf Proof of #1\/}}.}%
{\unskip\nobreak\hskip 2em plus 1fil\nobreak$\Box$
\parfillskip=0pt%
\endtrivlist}

\newcommand{\ignore}[1]{}

\theoremstyle{plain}
\newtheorem{claim}[theorem]{Claim}
\newcommand{\etal}{{\it et al.}\xspace}

\newlength{\myindent}
\newcommand{\bindent}{%
  \begingroup
  \setlength{\itemindent}{\myindent}
}
\newcommand{\eindent}{\endgroup}

\begin{document}

\maketitle

\begin{abstract}
A population protocol can be viewed as a
sequence of pairwise interactions of $n$ agents (nodes). During one
interaction, two agents selected uniformly at random update their
states by applying a specified deterministic transition function.  In a long run, the whole system
should stabilize at the correct output property. The main performance objectives in
designing population protocols are small number of states
per agent and fast stabilization time.

We present a fast population protocol for the exact-majority problem, which
uses $\Theta(\log n)$ states (per agent) and stabilizes in
$O(\log^{5/3} n)$ parallel time (i.e., in $O(n\log^{5/3} n)$ interactions)
in expectation and with high probability.
Alistarh~\etal~[SODA 2018] showed that any exact-majority protocol which stabilizes
in expected $O(n^{1-\epsilon})$ parallel time, for any constant $\epsilon > 0$, requires
$\Omega(\log n)$ states. They also showed an $O(\log^2 n)$-time protocol
with $O(\log n)$ states,
the currently fastest exact-majority protocol with polylogarithmic number of
states.
The standard design framework for majority protocols is based
on $O(\log n)$ phases and requires that all nodes are well synchronized within each phase,
leading naturally to upper bounds of the order of at least $\log^2 n$ because of $\Theta(\log n)$
synchronization time per phase.
%
%
We show how this framework can be tightened with {\em weak synchronization} to break
the $O(\log^2 n)$ upper bound of
previous protocols.
%
\end{abstract}

\newpage

\section{Introduction}

We consider population protocols~\cite{DBLP:journals/dc/AngluinADFP06} for
exact-majority voting. The underlying computation system consists of a
population of $n$ anonymous ({\it i.e.} identical) {\em agents}, or {\em
nodes}, and a {\em scheduler} which keeps selecting pairs of nodes for
interaction. A {\em population protocol\/} specifies how two nodes update their
states when they interact. The computation is a (perpetual) sequence of
interactions between pairs of nodes. The objective is for the whole system to
eventually stabilize in configurations which have the output property defined
by the considered problem. In the general case, the nodes can be connected
according to a specified graph $G=(V,E)$ and two nodes can interact only if
they are joined by an edge. Following the scenario considered in most
previous work on population protocols, we assume the complete communication
graph and the random uniform scheduler. That is, each pair of (distinct) nodes
has equal probability to be selected for interaction in any step and
each selection is independent of the previous interactions.

The model of population protocols was proposed in
Angluin~\etal~\cite{DBLP:journals/dc/AngluinADFP06} and has subsequently been
extensively studied to establish its computational power and to design
efficient solutions for fundamental tasks in distributed computing such as
various types of {\em consensus-reaching voting}. The survey from Aspnes
and Ruppert~\cite{AspnesR2009} includes examples of population
protocols, early computational results and variants of the model. The main
design objectives for population protocols are small number of states and fast
stabilization time. The original definition of the model assumes that the
agents are copies of the same finite-state automaton, so the number of states
(per node) is constant. This requirement has been later relaxed by allowing the
number of states to increase (slowly) with the population size, to study
trade-offs between the memory requirements and the run times.

The (two-opinion) {\em exact-majority voting} is one of the basic settings of
consensus voting~\cite{DBLP:conf/podc/AlistarhGV15,DBLP:journals/dc/AngluinADFP06,AngluinAE2008fast}.
Initially each node is in one of two distinct states $q_A$
and $q_B$, which represent two distinct opinions (or votes) $A$ and $B$, with
$a_0$ nodes holding opinion $A$ (starting in the state $q_A$) and $b_0$ nodes
holding opinion $B$. We assume that $a_0 \neq b_0$ and denote the initial
imbalance between the two opinions by $\ep = |a_0-b_0|/n \ge 1/n$. The desired
output property is that all nodes have the opinion of the initial majority. An
{\em exact\/} majority protocol should guarantee that the correct answer is
reached, even if the difference between $a_0$ and $b_0$ is only $1$
(cf.~\cite{DBLP:conf/podc/AlistarhGV15}). In
contrast, {\em approximate\/} majority would require correct answer only if the
initial imbalance is sufficiently large. In this paper, when we refer to
``majority'' (protocol, or voting, or problem) we always mean the
exact-majority notion.

We will now give further formalization of a population protocol and its time
complexity. Let $S$ denote the set of states, which can grow with the size~$n$
of the population (but keeping it low remains one of our objectives). Let
$q(v,t) \in S$ denote the state of a node $v\in V$ at step $t$ (that is, after
$t$ individual interactions). Two interacting nodes change their states
according to a common deterministic transition function $\delta: S\times S
\rightarrow S\times S$. A~population protocol has also an {\em output
function\/} $\g: S \rightarrow \G$, which is used to specify the desired output
property of the computation. For majority voting, $\g: S \rightarrow \, \{A,
B\}$, which means that a node in a state $q\in S$ assumes that $\g(q)$ is the
majority opinion. The system is in an (output) correct configuration at a step
$t$, if for each $v\in V$, $\g(q(v,t))$ is the initial majority opinion. We
consider undirected individual communications, that is, the two interacting
nodes are not designated as initiator and responder, so the transition
functions must be symmetric. Thus if $\delta(q', q'') = (r', r'')$,
then $\delta(q'', q') = (r'', r')$, implying, for example, that $\delta(q,q) =
(r,r)$.

We say that the system is in
a {\em stable configuration}, if no node will ever again change its output.
The computation continues (since it is perpetual) and nodes may continue updating their states,
but if a node changes from a state $q$ to another state $q'$,
then the output $\g(q')$ is the same as $\g(q)$.
Thus a majority protocol is in a correct stable configuration
if all nodes output the correct majority opinion and will do so in all possible subsequent configurations.
Two main types of output guarantee categorize population protocols as either
{\em always correct},
if they reach the correct stable configuration with probability $1$,
or {\em \whp correct}.
A protocol of the latter type
reaches a correct stable configuration \whp,\footnote{%
A property $P(n)$, {\em e.g.}\ that a given protocol reaches a stable correct configuration,
holds \whp\ (with high probability), if it holds with probability at least $1 - n^{-\a}$,
where constant $\a>0$ can be made arbitrarily large
by changing the constant parameters in $P(n)$
({\em e.g.}\ the constant parameters of a protocol).
}
allowing that with some low but positive probability
an incorrect stable configuration is reached or the computation does not stabilize at all.

The notion of the time complexity of population protocols
which has been used recently to derive lower bounds on the number of 
states~\cite{DBLP:conf/soda/AlistarhAEGR17,DBLP:conf/soda/AlistarhAG18},
and the notion which we use also in this paper, is the {\em stabilization time\/}
$T_S$ defined as the first {\em round\/} when the system enters a correct stable
configuration.%
\footnote{%
Some previous papers (e.g.~\cite{DBLP:conf/soda/AlistarhAEGR17,DBLP:conf/podc/BilkeCER17})
refer to this stabilization time as the convergence time.}
We follow the common convention of defining the {\em parallel time\/} as the number of
interactions divided by $n$. 
Equivalently, we group the
interactions in rounds of length $n$, called also ({\em parallel\/}) {\em steps},  and take the number of
rounds as the measure of time. In our analysis
we also use the term {\em period}, which we define as a sequence of
$n$ consecutive interactions, but not necessarily aligned with rounds.

\smallskip\smallskip

The main result of this paper is a majority
protocol with stabilization time $O(\log^{2-\delta} n)$ \whp\ and in
expectation, for some constant $\delta >0$ (here specifically
$\delta=1/3$), while using logarithmically
many states. According to~\cite{DBLP:conf/soda/AlistarhAG18} this 
number of states is asymptotically optimal for protocols with
$\E(T_S) = O(n^{1-\epsilon})$, and 
to the best of our knowledge this
is the first result that offers stabilization in time $O(\log^{2-\Omega(1)} n)$ with
poly-pogarithmic state space.


\subsection{Previous work on population protocols for the majority problem}

Draief and Vojnovi\'c~\cite{DBLP:conf/infocom/DraiefV10} and Mertzios
\etal~\cite{Mertzios-etal-ICALP2014} analyzed two similar four-state majority
protocols. Both protocols are based on the idea that the two opinions have {\em
weak\/} versions $a$ and $b$ in addition to the main {\em strong\/} versions
$A$ and~$B$. The strong opinions are viewed as tokens moving around the graph.
Initially each node $v$ has a strong opinion $A$ or $B$, and during the
computation it has always one of the opinions $a$, $b$, $A$ or $B$ (so is in
one of these four states). The strong opinions have dual purpose. Firstly, two
interacting opposite strong opinions cancel each other and change into weak
opinions. Such pairwise canceling ensures that the difference between the
number of strong opinions $A$ and $B$ does not change throughout the
computation (remaining equal to $a_0 - b_0$) and eventually
all strong opinions of the initial minority are canceled out.
 Secondly, the surviving strong
opinions keep moving around the graph, converting the weak opposite opinions.

\begin{sloppy}

Mertzios \etal~\cite{Mertzios-etal-ICALP2014} call their protocol the {\em
4-state ambassador protocol} (the strong opinions are ambassadors) and prove
the expected stabilization time $O(n^5)$ for any graph and $O((n \log n)/|a_0 -
b_0|)$ for the complete graph. Draief and
Vojnovi\'c~\cite{DBLP:conf/infocom/DraiefV10} call their 4-state protocol the
{\em binary interval consensus}, viewing it as a special case of the {\em
interval consensus} protocol of B{\'{e}}n{\'{e}}zit
\etal~\cite{DBLP:conf/icassp/BenezitTV09}, and analyze it in the
continuous-time model. For the uniform edge rates (the continuous setting which
is roughly equivalent to our setting of one random interaction per one time
unit) they show that the expected stabilization time for the complete graph is
at most $2n(\log n + 1)/|a_0 - b_0|$. They also derive bounds on the expected
stabilization time for cycles, stars and Erd{\H{o}}s-R\'enyi graphs.

\end{sloppy}

The appealing aspect of the four-state majority protocols is their simplicity
and the constant-size local memory, but the downside is polynomially slow
stabilization if the initial imbalance is small. The stabilization time
decreases if the initial imbalance increases, so the performance would be
improved if there was a way of boosting the initial imbalance. Alistarh
\etal~\cite{DBLP:conf/podc/AlistarhGV15} achieved such boosting by multiplying
all initial strong opinions by the integer parameter $r\ge 2$. The nodes keep the count of
the number of strong opinions they currently hold. When eventually all strong
opinions of the initial minority are canceled, $|a_0 - b_0|r$ strong opinions
of the initial majority remain in the system. This speeds up both the canceling
of strong opinions and the converting of weak opinions of the initial minority,
but the price is the increased number of states.
Refining this idea, Alistarh~\etal~\cite{DBLP:conf/soda/AlistarhAEGR17}
obtained a majority protocol which has the stabilization time $O(\log^3 n)$
\whp\ and in expectation and uses $O(\log^2 n)$ states.

A suite of polylogarithmic-time population protocols for various functions,
including the exact majority, was proposed by
Angluin~\etal~\cite{AngluinAE2008fast}. Their protocols are \whp correct and,
more significantly, require a unique leader to synchronize the progress of the
computation. Their majority protocol \whp reaches a correct stable
configuration within $O(\log^2 n)$ time (with the remaining low probability, it
either needs more time to reach the correct output or it stabilizes with an
incorrect output) and requires only a constant number of states, but the
presence of the leader node is crucial.

The protocols developed in~\cite{AngluinAE2008fast} introduced the idea of
alternating {\em cancellations\/} and {\em duplications}, which has been
frequently used in subsequent majority protocols and forms also
the basis of our new protocol. This idea has the following interpretation
within the framework of canceling strong opinions. The canceling stops when it
is guaranteed that \whp the number of remaining strong opinions is less than
$\d n$, for some small constant $\d < 1/2$. Now the remaining strong opinions
duplicate: if a node with a strong opinion interacts with a node which does not
hold a strong opinion, then both nodes get the same strong opinion. This
duplicating stops when it is guaranteed, again \whp, that all initial strong opinions
have been duplicated. One phase of (partial) cancellations followed by
(complete) duplications takes \whp $O(\log n)$ time, and $O(\log n)$
repetitions of this phase increases the difference between the numbers of
strong opinions $A$ and $B$ to $\Theta(n)$. With such large imbalance between
strong opinions, \whp within
additional $O(\log n)$ time
the minority opinion is completely eliminated and the majority opinion is
propagated to all nodes.

Bilke~\etal~\cite{DBLP:conf/podc/BilkeCER17} showed that the cancellation-duplication
framework from~\cite{AngluinAE2008fast} can be implemented without a leader if the agents
have enough states to count their interactions. They obtained a majority protocol which
has stabilization time $O(\log^2 n)$ \whp\ and in expectation, and uses $O(\log^2 n)$
states.
Berenbrink~\etal~\cite{DBLP:journals/corr/BerenbrinkFKMW16} considered population
protocols for the {\em plurality voting}, which generalizes the majority voting to $k \ge
2$ opinions. Using the methodology introduced earlier for load
balancing~\cite{DBLP:conf/focs/SauerwaldS12}, they generalized
the previous results on majority protocols
by working with multiple opinions and arbitrary graphs,
showing also only $O(\log n)$ time \whp\ for the case of complete graphs and
$k =2$.
Their protocol, however, requires a polynomial number of states and $\Omega(n/\polylog
n)$ initial advantage of the most common opinion to achieve $O(\polylog n)$ time.
Recently Alistarh~\etal~\cite{DBLP:conf/soda/AlistarhAG18} have shown that any majority
protocol which has expected stabilization time of $O(n^{1-\ep})$, where $\ep$ is any
positive constant, and satisfies technical conditions of monotonicity and output dominance,
requires $\Omega(\log n)$ states. They have also presented a protocol
which uses only $\Theta(\log n)$ states and has stabilization time $O(\log^2 n)$ \whp and
in expectation.

The lower and upper bounds shown in~Alistarh~\etal~\cite{DBLP:conf/soda/AlistarhAG18}
raised the following questions.
Can exact majority be computed in poly-logarithmic time with $o(\log n)$ states, if 
the time complexity is measured in some other natural and relevant way than the time till (correct) stabilization?
Can exact majority be computed in $o(\log^2 n)$ time  with poly-logarithmic states?
(The protocol in~\cite{DBLP:conf/soda/AlistarhAG18} and all earlier exact majority protocols
which use poly-logarithmic number of states have time complexity at least of the order of $\log^2 n$.)
For a random (infinite) sequence $\omega$ of interaction pairs, let $T_C = T_C(\omega)$ denote 
the {\em convergence time}, defined as the first round when (at some interaction during this round)  
the system enters a correct configuration
(all nodes correctly output the majority opinion) and remains in correct configurations
in all subsequent interactions (of this sequence  $\omega$).
Clearly $T_C \le T_S$, since reaching a correct stable configuration 
implies that whatever the future interactions may be, the system will always remain 
in correct configurations.

Very recently
Kosowski and Uzna{\'n}ski~\cite{2018arXiv180206872K} and Berenbrink \etal~\cite{All-PODC2018submission}
have shown that the convergence time $T_C$ can be poly-logarithmic
while using $o(\log n)$ states.%
In~\cite{2018arXiv180206872K} the authors design a programming framework
and accompanying compilation schemes that provide a simple way of achieving protocols
(including majority) which are \whp correct, use $O(1)$ states
and converge in expected poly-logarithmic  time. They can make their protocols
always-correct at the expense of having to use $O(\log\log n)$ states per
node, while keeping poly-logarithmic time, or increasing time to 
$O(n^\epsilon)$, while keeping a constant bound on the number of states.
In~\cite{All-PODC2018submission} the authors design an always-correct 
majority protocol which converges \whp in 
$O(\log^2 n/{\log s})$ time and uses $\Theta(s + \log\log n)$ states
and an always-correct 
majority protocol which stabilizes \whp in 
$O(\log^2 n/{\log s})$ time and uses $O(s \cdot \log n / {\log s})$ states,
where parameter $s\in [2,n]$.

The recent population protocols for majority voting
often use similar technical tools (mainly the same efficient constructions of {\em phase clocks})
as protocols for another fundamental problem of {\em leader election}.
There are, however, notable differences in computational difficulty of both problems, so
advances in one problem do not readily imply progress with the other problem.
For example, leader election admits always-correct protocols with
poly-logarithmically fast stabilization and
only $\Theta(\log\log n)$ states (the lower bound here is only $\Omega(\log\log
n)$~\cite{DBLP:conf/soda/AlistarhAEGR17}). There are some general ideas, recently
explored in~\cite{DBLP:journals/corr/abs-1802-06867}, which indicate that in leader
election expected run times of order significantly better than $\log^2 n$ can be
achieved  (though the \whp time would remain $\Theta(\log^2 n)$).
Those ideas, however, are specific for leader election and not applicable to majority voting.

\subsection{Our contributions}

We present a majority population protocol with stabilization time $O(\log^{5/3} n)$ \whp
and in expectation, using asymptotically optimal $O(\log n)$ states.
This is the first
state-space optimal protocol for majority with stabilization time $O(\log^{2-\Omega(1)})$.
In fact, to the best of our knowledge, there is no other majority protocol with
$O(\polylog n)$ states and time $O(\log^{2-\Omega(1)})$,
even for the weaker notions of the convergence time or \whp-correctness.


All known fast majority population protocols using a polylogarithmic number of states are
based in some way on the idea of a sequence of $\Omega(\log n)$
canceling-duplicating
(or canceling-doubling) phases, each of length $\Omega(\log n)$ (first introduced
in~\cite{AngluinAE2008fast}),
synchronizing the nodes
across phase boundaries.
In our
new protocol we still use the canceling-doubling framework
(as explained in Section \ref{generalidea})
but with \emph{shorter phases}
of length $\log^{1-\Omega(1)} n$ each, at the expense of loosing the
synchronization. We note that all existing protocols known to us working within the
canceling-doubling framework
cease to function properly with such short
phases.
Not only can we no longer guarantee a synchronized transition across phase
boundaries (and in order to obtain the correct answer one must not allow opposite opinions
from different phases to cancel each other), but we do not even have the guarantee
that every node will be activated at all
during a phase (in fact, we know some will not). The existing protocols require each node
to be activated at least once (actually at least logarithmically many times) during
each phase.
Our main technical contributions are mechanisms to deal with nodes advancing too
slowly or too quickly through the short phases, that is, nodes which are not in sync with the bulk.
We believe that some
algorithmic and analytical ideas used for this may be of independent interest.

\section{Exact majority: the general idea of canceling-doubling phases, and a protocol with $O(\log^2 n)$ time and $\Theta(\log^{2} n)$ states}
\label{generalidea}

We view the $A/B$ votes as tokens, which can have different ages and values
(magnitudes). Initially each node has one token of type $A$ or $B$, with age
$0$ and value $1$. Throughout the computation, each node either has one token
or is {\em empty}. In the following we say that two tokens {\em meet} if
their corresponding nodes interact.

\begin{itemize}\itemsep0pt
\item When two opposite tokens (one  $A$ and the other $B$) of the same value meet,
then
they
cancel each other and the nodes become empty. Such an interaction is called
canceling.
\item When a token of type $X\in \{A,B\}$ and age $g$ interacts with an empty node, then this token
splits into two tokens, each of type $X$, age $g+1$ and half the value, and each of the two involved nodes
takes one token.
We refer to such splitting of a token also as {\em duplicating\/} or {\em doubling}.
\end{itemize}

Thus the age $g$ of a token is equal to the number of times it has undergone
splitting; its value is equal to $1/2^g$. Note that any sequence of canceling
and splitting interactions preserves the difference between the sum of the
values of all $A$ and $B$ tokens. This  difference is always equal to the initial
imbalance. The primary objective is to eliminate all minority tokens.
When only majority tokens are left in the system, the majority opinion can be
propagated to all nodes \whp within additional $O(n\log n)$ interactions via a
broadcast process. The final standard process of propagating the outcome will be omitted
from our descriptions and analysis. That is, from now on we assume that the
objective is to eliminate the minority tokens.

We first, in this section, describe the $O(\log^2 n)$-step $O(\log^2 n)$-state \ExactMajority
protocol presented in~\cite{DBLP:conf/podc/BilkeCER17}. Then we
propose two new protocols, both with a runtime of $O(\log^{5/3} n)$ steps:
\FastExactMajorityA with $\Theta(\log^2 n)$ states
(described and analyzed in Sections~\ref{Sec:protocol} and~\ref{analysis})
and \FastExactMajorityB
with $\Theta(\log n)$ states (outlined in Section~\ref{Sec-FastOptStates}).
Further details of our protocols, including pseudocodes and detailed proofs, are given in
\ifthenelse{\isundefined{\ConferenceVersion}}{Appendix}{the full version of the paper~\cite{fullversion}}.

The structure of the $O(\log^2 n)$-step
\ExactMajority protocol will provide a useful reference in explanations of the computation and the
analysis of the faster protocols.
From the node's local point of view, the computation of the \ExactMajority
protocol consists of at most $\log n +2$ phases and each phase consists of
at most $C\log n$ interactions, where $C$ is a suitably large constant. Each
node keeps track of the number of phases and steps (interactions) within the
current phase, and maintains further information which indicates the
progress of computation. More precisely, each node $v$ keeps
the following data, which require $\Theta(\log^2 n)$ states.

\begin{itemize}
\item
$v.\vtoken \in \{A, B, \notoken\}$ --
the type of token held by $v$. If $v.\token = \notoken$ then the node is empty.
\item
$v.\phase\in\{0,1,2,\ldots, \log n + 2\}$ -- the counter of phases.
\item
$v.\phasestep\in\{0,1,2,\ldots,(C\log n) - 1\}$ -- the counter of  steps in the current phase.
\item
Boolean flags, which are initially false and indicate the following status when set to $\true$:
\begin{itemize}
\item
$v.\doubled$ -- $v$ has a token which has already doubled in the current phase;
\item
$v.\done$ -- the node has made the decision on the final output;
\item
$v.\fail$ -- the protocol has failed because of some inconsistencies.
\end{itemize}
\end{itemize}

If a node $v$ is in neither of the two special states $\done$ and $\fail$, then
we say that $v$ is in a {\em normal state}:
$v.\normal \;\; \equiv \;\; \neg(v.\done \vee v.\fail).$ A node $v$ is in
Phase $i$ if $v.\phase = i$. If $v$ is in Phase $i$ and is not empty, then the
age of the token at $v$ is either $i$ if not $v.\doubled$ (the token has not
doubled yet in this phase) or $i+1$ if $v.\doubled$. Thus the phase of a token
(the phase of the node where the token is) and the flag $\doubled$ indicate the
age of this token. Throughout the whole computation, the pair
$(v.\phase,v.\phasestep)$ can be regarded as the (combined) interaction counter
$v.\Time\in \{0,1,2,\ldots,2C\log^2 n)\}$ of node $v$. This counter is
incremented by $1$ at the end of each interaction. Thus, for example, if
$v.\phasestep$ is equal to $0$ after such an increment, then node $v$ has
just completed a phase.

Each phase
is divided into five parts defined below, where $c$
is a constant discussed later.

\begin{itemize}\itemsep0pt
\item The second part is the {\em canceling stage\/}
and the fourth part is the {\em doubling stage},
each consisting of $((C- 3c)/2)\log n$ steps.
If two interacting nodes are in the canceling stage of the same phase and have opposite tokens,
then the tokens cancel out.
If two interacting nodes are in the doubling  stage of the same phase,
one of them has a token which has not doubled yet in this phase and the other  is empty,
then this is a doubling interaction.
\item The beginning, the middle and the final parts of a phase
are buffer zones, consisting of
$c\log n$ steps each. The purpose of these parts is to ensure that
the nodes progress through the current phase in a synchronized way.
\end{itemize}

If nodes were simply incrementing their step counters by $1$ at each
interaction, then those counters would start diverging too much for the
canceling-doubling process to work correctly. An important aspect of the
$\ExactMajority$ protocol, as well as our new faster protocols, is the
following mechanism for keeping the nodes sufficiently synchronized. When two
interacting nodes are in different phases, then the node in the lower phase
jumps up to (that is, sets its step counter to) the beginning of the next
phase. The $\ExactMajority$ protocol relies on this synchronization mechanism
in the high probability case when
all nodes
are in two adjacent parts of a phase (that is, either in two consecutive parts
of the same phase, or in the final part of one phase and the beginning part of
the next phase.)
In this case the process of pulling
all nodes up to the next phase follows the pattern of {\em broadcast}. The
node, or nodes, which have reached the beginning of the next phase by way
of normal one-step increments broadcast the message ``\emph{if you are not yet
in the same phase as I am, then jump up to the next phase.}'' By the time the
broadcast is completed (that is, by the time when the message has reached all
nodes), all nodes are together in the next phase. It can be shown that there is a constant
$\beta_0$ such that \whp the broadcast
completes in $\beta_0 n \log n$ random pairwise interactions (see, for
example~\cite{AngluinAE2008fast};
other papers may refer to this process  as {\em epidemic spreading} or {\em rumor spreading}).

The constant $c$ in the definition of the parts of a phase
is suitably smaller than the constant $C$, but sufficiently
large to guarantee the following two conditions: ({\it a\/}) the {\em
broadcast\/} from a given node to all other nodes completes \whp within
$(c/5)n\log n$ interactions; and ({\it b\/}) for a sequence of $(C/2) n\log n$
consecutive interactions, \whp for each node $v$ and each $0 < t \le (C/2)n\log
n$, the number of times $v$ is selected for interaction within the first $t$
interactions differs from the expectation (which is equal to $2t/n$) by at most
$(c/5)\log n$. Condition~({\it a\/}) is used when the nodes reaching the
end of the current phase $i$ initiate broadcast to ``pull up'' the nodes
lagging behind. Condition~({\it a\/}) implies that after $(c/5)n\log n$
interactions, \whp all nodes are in the next phase. Using Condition~({\it b\/})
with $t = (c/5)n\log n$, we can also claim that \whp at this point all nodes
are within the first $(3/5)c\log n$ steps of the next phase (all nodes are in
the next phase and no node interacted more than the expected $(2/5)c\log n$
plus $(1/5)c\log n$ times). Finally Condition~({\it b\/}) applied to all
$(c/5)n\log n \le t \le (C/2)n\log n$ implies that \whp\ the differences
between the individual counts of node interactions do not diverge by more than
$c\log n$ throughout this phase. We set $c = C^{3/4}$ and take $C$ large enough
so that $c \le C/9$ (to have at least $3c\log n$ steps in the canceling and
doubling stages) and both Conditions~({\it a\/}) and~({\it b\/}) hold. This way
we achieve the following synchronized progress of nodes through a phase: \whp
all nodes are in the same part of the same phase before they start moving on to
the next part. Moreover, also \whp, for each canceling or doubling stage there
is a sequence of consecutive $\Theta(n\log n)$ interactions when all nodes
remain in this stage and each of them is involved in at least $c\log n$
interactions.

Thus throughout the computation of the \ExactMajority protocol, \whp all nodes
are in two adjacent parts of a phase. In particular, \whp\ the canceling and doubling activities of
the nodes are separated. This separation ensures that the cancellation of
tokens creates a sufficient number of empty nodes to accommodate new tokens
generated by token splitting in the subsequent doubling stage.
If two
interacting nodes are not in the same or adjacent parts of a phase (a low, but
positive, probability), then their local times (step counters) are considered
inconsistent and both nodes enter the special $\fail$ state.
\ifthenelse{\isundefined{\ConferenceVersion}}{%
The details of the \ExactMajority protocol are given in pseudocode
in Algorithms~\ref{BasicAlgo} and~\ref{BasicNextStepCode}.
}{%
The details of the \ExactMajority protocol are given in pseudocode in
the full version~\cite{fullversion}.
}

From a global point of view, \whp\ each new phase $p$ starts with all nodes
in normal states in the beginning of this phase. We say that this phase
completes successfully if all nodes are in normal states in the beginning part
of the next phase $p+1$. At this point all tokens have the same value
$1/2^{p+1}$, and the difference between the numbers of opposite tokens is equal
to $2^{p+1} |a_0 - b_0|$. The computation \whp\ keeps successfully completing
consecutive phases, each phase halving the value of tokens and doubling the
difference between $A$ tokens and $B$ tokens, until the {\em critical phase\/}
$p_c$, which is the first phase $0 \le p_c \le \log n -1$ when the difference
between the numbers of opposite tokens is
\begin{equation}\label{critical-phase}
 2^{p_c} |a_0 - b_0| > \frac{n}{3}.
\end{equation}
The significance of the critical phase is that the large difference between the
numbers of opposite tokens means that \whp\ all minority tokens will be
eliminated in this phase, if they have not been eliminated yet in previous
phases. More specifically, at the end of phase $p_c$, \whp\ only tokens of the
majority opinion are left and each of these tokens has value either
$1/2^{p_c+1}$, if the token has split in this phase, or $1/2^{p_c}$, otherwise.
If at least one token has value $1/2^{p_c}$, then this token has failed to
doubled during this phase and assumes that the computation has completed.
Such a node enters the $\done$ state and broadcasts its (majority) opinion to
all other nodes. In this case phase $p_c$ is the {\em final phase}.

If at the end of the critical phase all tokens have value $1/2^{p_c+1}$, then no
node knows yet that all minority tokens have been eliminated, so the
computation proceeds to the next phase $p_c+1$. Phase $p_c+1$ will be the final
phase, since it will start with more than $(2/3)n$ tokens
and all of them of the same type, so at least one
token will fail to double and will assume that the computation has completed and will enter the $\done$ state.
The condition that a token has failed to double is taken as indication that
\whp all
tokens of opposite type have been eliminated.
Some tokens may still double in the final phase and enter the next phase (receiving later
the message that the computation has completed)
but \whp no node reaches the end of phase $p_c +2 \le \log n + 1$.
Thus the $\done$ state is reached \whp within $O(\log^2 n)$ parallel time.

The computation may fail \wlp\footnote{%
\wlp\ --  {\em with low probability} -- means that the opposite event happens \whp}
when the step counters of two interacting nodes are not consistent, or a node
reaches phase $\log n +2$, or two nodes enter the $\done$ state with opposite
type tokens. Whenever a node realizes that any of these low probability events
has occurred, it enters the $\fail$ state and broadcasts this state to all
other nodes.
\ifthenelse{\isundefined{\ConferenceVersion}}{%
(The standard broadcast of $\done$ and $\fail$ states is not
included in the pseudocodes.)}{}

It is shown in~\cite{DBLP:conf/podc/BilkeCER17} that
the \ExactMajority protocol stabilizes, either in
the correct all-$\done$ configuration or in the all-$\fail$ configuration,
within $O(\log^2 n)$ time \whp and in expectation. The standard technique of
combining a fast protocol, which \wlp may fail, with a slow but always-correct backup
protocol gives an {\em extended\/} \ExactMajority protocol, which requires
$\Theta(\log^2 n)$ states per node and computes the exact majority within
$O(\log^2 n)$ time \whp and in expectation. For the slow always-correct protocol take
the four-state majority protocol, run both the fast and the slow
protocols in parallel and make the nodes in the $\fail$ state adopt the outcome
of the slow protocol. The slow protocol runs in expected polynomial time, say
in $O(n^\alpha)$ time, but its outcome is used only with low probability of
$O(n^{-\alpha})$, so it contributes only $O(1)$ to the overall expected time.

We omit the details of using a slow backup protocol (see, for example,
\cite{DBLP:conf/soda/AlistarhAG18,DBLP:conf/podc/BilkeCER17}), and assume that
the objective of a canceling-doubling protocol is to use a small number of
states~$s$, to compute the majority quickly \whp, say within a time bound
$T'(n)$, and to have also low expected time of reaching the
correct all-$\done$ configuration or the all-$\fail$ configuration, say within
a bound $T''(n)$. If the bounds $T'(n)$ and $T''(n)$ are of the same order
$O(T(n))$, then we get a corollary that the majority can be computed with
$O(s)$ states in $O(T(n))$ time \whp and in expectation.
%
%
%
%
%
%


\section{Exact majority in $O(\log^{5/3} n)$ time with $\Theta(\log^{2} n)$ states}
\label{Sec:protocol}

To improve on the $O(\log^2 n)$ time of the \ExactMajority protocol, we shorten
the length of a phase to $\Theta(\log^{1-a} n)$, where $a = 1/3$. The new
\FastExactMajorityA protocol runs in $O(\log^{1-a} n) \times O(\log n) =
O(\log^{5/3} n)$ time and requires $\Theta(\log^{2} n)$ states per node. We
will show in Section~\ref{Sec-FastOptStates} that the number of states can be
reduced to asymptotically optimal $\Theta(\log n)$. We keep the term $a$ in the description and the
analysis of our fast majority protocols to simplify notation and to make it
easier to trace where a larger value of $a$ would break the proofs.

Phases of sub-logarithmic length are too short to ensure that \whp\ all tokens
progress through the phases synchronously and keep up with required canceling
and doubling, as they did in the \ExactMajority protocol. In the
\FastExactMajorityA protocol, we have a small but \whp positive number of {\em
out-of-sync\/} tokens, which move to the next phase either too early or too
late (with respect to the expectation) or simply do not succeed with splitting within
a short phase. Such tokens stop contributing to the regular dynamics of
canceling and doubling.
The general idea of our solution is to group $\log^a n$ consecutive phases (a total of
$\Theta(\log n)$ steps) into an {\em epoch}, to attach further $\Theta(\log n)$
steps at the end of each epoch to enable the out-of-sync tokens to reach the
age required at the end of this epoch, and to synchronize all nodes by the
broadcast process at the boundaries of epochs. When analyzing the progress of
tokens through the phases of the same epoch, we  consider  separately the
tokens which remain synchronized and the out-of-sync tokens.

We now proceed to the details of the \FastExactMajorityA protocol. Each epoch
consists of $2C\log n$ steps, where $C$ is a suitably large constant,
and is divided into two equal-length
parts.
 The first part is a sequence of
$\log^a n$ canceling-doubling phases, each of length $C\log^{1-a} n$. The
purpose of the second part is to give sufficient time to out-of-sync tokens so
that \whp\ they all complete all splitting required for this epoch.
Each node $v$ maintains the following data, which can be stored using $\Theta(\log^2 n)$ states.
For simplicity of notation, we assume that expressions like $\log^a n$ and
$C\log^{1-a} n$ have integer values if they refer to an index (or a number) of
phases or steps.

\begin{itemize}
\item
$v.\vtoken \in \{A, B, \emptyset\}$ -- type of token held by $v$.
\item
$v.\epoch\in\{0,1,\ldots,\log^{1-a}n + 2\}$ - the counter of epochs.
\item
$v.\ageinepoch\in \{0,1,\ldots,\log^{a}n\}$ -- the age of the token at $v$ (if $v$ has a token)
with respect to the beginning of the current epoch.
If $v.\token$ is $A$ or $B$, then the age of this token is
$g = v.\epoch \cdot \log^{a}n + v.\ageinepoch$ and
the value of this token is $1/2^g$.
\item
$v.\epochpart\in \{ 0,1\}$ -- each epoch consists of two parts, each part has $C\log n $ steps.
The first part, when $v.\epochpart=0$, is divided into $\log^a n$ canceling-doubling phases.
\item
$v.\phase\in\{0,1,\ldots,(\log^{a}n) - 1\}$ -- counter of phases in the first part of the
current epoch.
\item
$v.\phasestep\in\{0,1,\ldots,(C\log^{1-a}n) - 1\}$ -- counter of steps (interactions)
in the current phase.
\item
Boolean flags indicating the status of the node, all set initially to $\false$:
\begin{itemize}
\item
$v.\doubled$,
$v.\done$,
$v.\fail$ -- as in the \ExactMajority protocol;
\item
$v.\outofsync$ -- $v$ has a token which no longer follows the expected progress through the phases
of the current epoch;
\item
$v.\additionalepoch$ -- the computation is in
the additional epoch of $3\log^a n$ phases,
with each of these phases consisting  now of $\Theta(\log n)$ steps.
\end{itemize}
\end{itemize}


\begin{sloppy}

We say that a node $v$ is in epoch $j$ if $v.\epoch = j$, and in phase $i$ (of
the current epoch) if \mbox{$v.\phase = i$}. We view the triplet
$(v.\epochpart, v.\phase, v.\phasestep)$ as the (combined) counter
$v.\epochstep\in\{0,1,2,\ldots,(2C\log n) - 1\}$ of steps in the current epoch,
and the pair $(v.\epoch,v.\epochstep)$ as the counter $v.\Time\in
\{0,1,2,\ldots, 2C\log^{2-a} n)+O(\log n)\}$ of the steps of the whole protocol.
If a node $v$ is not in any of the special states $\outofsync,
\additionalepoch, \done$ or $\fail$, then we say that $v$ is in a {\em normal
state}:
\[ v.\normal \;\; \equiv \;\; \neg(v.\outofsync \vee v.\additionalepoch \vee v.\done \vee v.\fail).
\]
A normal token is a token in a normal node.
Each phase is split evenly into the canceling stage (the first $(C/2)\log^{1-a}
n$ steps of the phase) and the doubling stage (the remaining $(C/2)\log^{1-a}
n$ steps).

\end{sloppy}

The vast majority of the tokens are normal tokens progressing through the
phases of the current epoch in a synchronized fashion. These tokens are at the
same time in the beginning part of the same phase $j$ and have the same age
$j$ (\wrt the end of the epoch).
They first try to cancel out with tokens of the same age but opposite type
during the canceling stage, and if they survive, then they split during the
subsequent doubling stage. At some later time most of the tokens will still be
normal, but in the beginning part of the next phase $j+1$ and having age $j+1$.
Thus the age of a normal token (w.r.t.\ the beginning of the current epoch) is
equal to its phase, if the token has not split yet in this phase, or to its
phase plus $1$, if the token has split (this is recorded by setting the flag
$\doubled$).

As in the \ExactMajority protocol, we separate the canceling and the doubling
activities to ensure that the canceling of tokens creates first a sufficient
number of empty nodes to accommodate the new tokens obtained later from splitting.
Unlike in the \ExactMajority protocol, the
\FastExactMajorityA protocol does not have the buffer zones within a phase.
Such zones would not be helpful in the context of shorter
sublogarithmic phases when anyway we cannot guarantee that \whp\ {\em all\/}
nodes progress through a phase in a synchronized way.

A token which has failed to split in one of the phases of the current epoch
becomes an out-of-sync token (the $\outofsync$ flag is set). Such a token no
longer follows the regular canceling-doubling phases of the epoch, but instead
tries cascading splitting to break up into tokens of age $\log^{a} n$ (relative
to the beginning of the epoch) as expected by the end of this epoch. An
out-of-sync token does not attempt canceling out because there would be only
relatively few opposite tokens of the same value, so small chance to meet them
(too small to make a difference in the analysis). The tokens obtained from
splitting out-of-sync tokens inherit the out-of-sync status. A token drops the
out-of-sync status if it is in the second part of the epoch and has reached the
age $\log^{a} n$. (Alternatively, out-of-sync tokens could switch back to the
normal status as soon as their age coincides again with their phase, but this
would complicate the analysis.) An {\em out-of-sync node\/} is a node with an
out-of-sync token. While each normal node and token is in a specific phase of
the first part of an epoch or is in the second part of an epoch, the out-of-sync nodes (tokens) belong to
an epoch but not to any specific phase. The objective for a
normal token is to split into two halves in each phase of the current epoch (if
it survives canceling). The objective of an out-of-sync token is to keep
splitting in the current epoch (disregarding the boundaries of phases)
until it breaks into tokens expected at the end of this epoch.

We show in our analysis that \whp there are only $O(n/2^{\Theta(\log^a n)})$
out-of-sync tokens in one epoch. \Whp all out-of-sync tokens in the current epoch reach the age $\log^{a} n$
(w.r.t.\ the beginning of the epoch) by the mid point of the second part of the
epoch (that is, by the step $(3/2)C\log n$ of the epoch), for each epoch before
the {\em final epoch\/} $j_f$. In the final epoch at least one out-of-sync token completes the
epoch without reaching the required age.

When the system completes the final epoch, the task of determining the
majority opinion is not fully achieved yet. In contrast to the \ExactMajority
protocol where on the completion of the final phase \whp only majority tokens
are left, in the \FastExactMajorityA protocol there may still be a small number
of minority tokens at the end of the final epoch, so some further work is
needed. A node which has failed to reach the required age by the end of the
current epoch, identifying that way that this is the final epoch, enters the
$\additionalepoch$ state and propagates this state through the system to
trigger an {\em additional epoch\/} of $\Theta(\log^{a} n)$ phases. More
precisely, the additional epoch consists of at most $3\log^{a} n$ phases
corresponding to epochs $j_f - 1$ (if $j_f > 0$), $j_f$ and $j_f+1$, and each phase
has now $\Theta(\log n)$ steps. \Whp these phases include the critical
phase $p_c$ and the phase $p_c+1$, defined by~\eqref{critical-phase}. The computation of the additional epoch is
as in the $\ExactMajority$ protocol, taking $O(\log^{1+a} n)$ time to
reach the correct all-$\done$ configuration \whp or the all-$\fail$
configuration \wlp

Two interacting nodes first check the consistency of their $\Time$ counters
(the counters of interactions) and switch to $\fail$ states, if the difference
between the counters is greater than $(1/4)C\log n$. If the counters are
consistent but the nodes are in different epochs (so one in the end of an
epoch, while the other in the beginning of the next epoch), then the node in
the lower epoch jumps up to the beginning of the next epoch. This is the
synchronization mechanism at the boundaries of epochs, analogous to the
synchronization by broadcast at the boundaries of phases in the \ExactMajority
protocol. In the \FastExactMajorityA protocol, however, it is not possible to synchronize
the nodes at the boundaries of (short) phases.

\ifthenelse{\isundefined{\ConferenceVersion}}{%
For details of the \FastExactMajorityA protocol, see the pseudocodes given in
Algorithms~\ref{AlgoNormalProgress}--\ref{NextStepCodeOutofsync} in the Appendix.
The pseudocodes do not include the details of the additional epoch, since this final part of the computation
follows closely the \ExactMajority protocol.
To enable the initialization of the additional epoch, the nodes keep track of the tokens they
have had at the end of the two previous epochs.
(All nodes have to know their tokens from the beginning of epoch $j_f - 1$,
but there may be nodes which have already progressed to epoch $j_f +1$ when they are notified
that epoch $j_f$ is the final one.)
The additional epoch
does not need additional states since it can (re-)use the existing states.
}{%
For details of the \FastExactMajorityA protocol we refer the reader to the full version~\cite{fullversion}.
}

\section{Analysis of the \FastExactMajorityA protocol}
\label{analysis}

Ideally, we would like that \whp\ all tokens progress through the phases of the
current epoch in a synchronized way, that is, all tokens are roughly in the
same part of the same phase, as in the \ExactMajority protocol. This would mean
that \whp\ at some (global) time all nodes are in the beginning part of
the same phase, ensuring that all tokens have the same value $x$, and at some
later point all nodes are in the end part of this phase and all surviving
tokens have value $x/2$. This ideal behavior is achieved by the \ExactMajority
protocol at the cost of having $\Theta(\log n)$-step phases.
As discussed in Section~\ref{generalidea}, the logarithmic length of a phase
gives also sufficient time to synchronize \whp\ the local times of all nodes at
the end of a phase so that they all end up together in the beginning part of
the next phase.

Now, with phases having only $\Theta(\log^{1-a} n)$ steps, we face the
following two difficulties in the analysis. Firstly, while a good number of
tokens split during such a shorter phase, \whp\ there are also some tokens
which do not split. Secondly, phases of length $o(\log n)$ are too short to
keep the local times of the nodes synchronized. We can show again that a good
number of nodes proceed in synchronized manner, but \whp\ there are nodes
falling behind or rushing ahead and our analysis has to account for them.


Counting the phases across the epochs, we define the critical phase $p_c$ as
in~\eqref{critical-phase}. Similarly as in the $O(\log^2 n)$-time
\ExactMajority protocol, the computation proceeds through the phases moving
from epoch to epoch until the critical phase $p_c$. Then the computation gets
stuck on this phase or on the next phase $p_c+1$. Some tokens do not split in
that final phase or in any subsequent phase of the current epoch because there
are not enough empty nodes to accommodate new tokens. Almost all minority
tokens have been eliminated, so the creation of empty nodes by cancellations of
opposite tokens has all but stopped. This is the final epoch $j_f$ and the
nodes which do not split to the value required at the end of this epoch trigger
the additional epoch of $O(\log^a n)$ phases, each having $\Theta(\log n)$
steps. The additional epoch is needed because we do not have a high-probability
guarantee that all minority tokens are eliminated by the end of the final
epoch. The small number of remaining minority tokens may have various values
which are inconsistent with the values of the majority tokens, so further
cancellations of tokens might not be possible. The additional epoch includes
the phases of the three consecutive epochs $j_f-1, j_f$ and $j_f+1$ to ensure that \whp both phases $p_c$
and $p_c+1$ are included. Phase $p_c$ can be as early as the last phase in
epoch $j_f - 1$ and phase $p_c +1$ can be as late as the first phase in epoch
$j_f +1$.

The following conditions
describe the regular configuration of the whole system at the beginning of
epoch $j$, and the corresponding Lemma~\ref{Lem:epochinv}
summarizes the progress of the computation through this epoch.
Recall that the \FastExactMajorityA protocol is parameterized by a
suitably large constant $C > 1$ and our analysis refers also to another smaller
constant $c = C^{3/4}$.
We refer to the first (resp.\ the last) $c\log^{1-a} n$ steps of a
phase or a stage as the beginning (resp.\ the end) part of this phase or stage.
The (global) time steps count the number of interactions of the whole system.

\smallskip

\noindent $\EpochInv(j):$
\begin{enumerate}
\item\label{EpochInvNormalNodes}
At least $n(1 - 1/2^{3\log^{a}n})$ nodes are in normal states, are in epoch $j$, and
their $\epochstep$ counters are at most $c\log^{a}n$.
\item
For each remaining node $u$,
\begin{enumerate}
\item\label{EpochInvFastNodes}
$u$ is in a normal state in epoch $j-1$ and $u.\epochstep \ge (3 /2)C\log n$ (that is, $u$ is in the last quarter of
epoch $j-1$), or
\item\label{EpochInvSlowNodes}
$u$ is in a normal or out-of-sync state in epoch $j$ and $u.\epochstep \le 4c\log n$.
\end{enumerate}
\end{enumerate}

\begin{lemma}
\label{Lem:epochinv}
Consider an arbitrary epoch $j \ge 0$ such that phase $p_c$ belongs to an epoch $j' \ge j$
and assume  that at some (global) step $t$ the condition $\EpochInv(j)$ holds.
\begin{enumerate}
\item
If phase $p_c$ does not belong to epoch $j$ (that is, phase $p_c$ is in a later epoch $j' > j$), then
\whp\
there is a step $\tilde{t} \le t + 2Cn\log n$ when
the condition $\EpochInv(j+1)$ holds.
\item
If both phases $p_c$ and $p_c+1$ belong to epoch $j$, then
\whp\
there is a step $\tilde{t} \le t + 2Cn\log n$ when
\begin{enumerate}
\item[$(*)$]
a nodes is completing epoch $j$ and enters the $\additionalepoch$ state
(because it has a token which has not split to the value required
at the end of this epoch); and

all other nodes are in normal or out-of-sync
states in the second part of epoch $j$ or the first part of epoch $j+1$.
\end{enumerate}
\item
Otherwise, that is, if phase $p_c$ is the last phase in epoch $j$
(and $p_c+1$ is the first phase in epoch $j+1$),
then \whp
either there is a step $\tilde{t} \le t + 2Cn\log n$ when
the above condition $(*)$ for the end of epoch $j$ holds, or
all nodes eventually progress to epoch $j+1$
and there is a step $\hat{t} \le t + 3Cn\log n$ when the condition analogous to $(*)$ but for the end
of epoch $j+1$ holds.
\end{enumerate}
\end{lemma}


The condition $\PhaseInvA(j,i)$ given below
describes the regular configuration of the whole system at the beginning of
phase $0 \le i \le \log^a n$ in epoch $j\ge 0$. We note that the last phase in
an epoch is phase $\log^a n -1$ and the condition $\PhaseInvA(j,\log^a n)$
refers in fact to the beginning of the second part of the epoch. A normal token
in the beginning of phase $i$ in epoch $j$ has (absolute) value $1/2^{j\log^{a}
n + i}$ and relative values $1$, $2$, $1/2^{i}$ and $2^{\log^{a} n - i}$ \wrt
(the beginning of) this phase, the end of this phase, the beginning of this
epoch and the end of this epoch, respectively. It may also be helpful to recall
that for a given node $v$, phase $i$ starts at $v$'s epoch step $Ci\log^{1-a}
n$. Observe that $\EpochInv(j)$ implies $\PhaseInvA(j,0)$.
\smallskip

\noindent$\PhaseInvA(j,i):$
\begin{enumerate}
\item \label{hjjx90sa}
The set $W$ of nodes which are normal and in the beginning part
of phase~$i$ in epoch~$j$ has size
at least $n(1 - (i+1)/2^{2\log^{a}n})$.
That is, a node $v$ is in $W$ if, and only if, $v.\normal$ is true,
$v.\phasestep \le c\log^{1-a} n$, $v.\epoch = j$, and either
$v.\epochpart = 0$ and $v.\phase = i$, if $ i < \log^a n$,
or $v.\epochpart = 1$ and $v.\phase = 0$, if $ i = \log^a n$.
\item\label{hjjsa}
Let $U=  V \setminus W$ denote the set of the remaining nodes.
\begin{enumerate}
\item\label{k892a}
For each $u \in U$:

$u$ is a normal node in epoch $j-1$, $u.\epochstep \ge (3/2)C\log n$
and $i < (c/C)\log^{a} n$; or
%
$u$ is in a normal or out-of-sync state in epoch $j$ and $|u.\epochstep - Ci\log^{1-a} n| \le 4c\log n$.
\item\label{hk892a}
The total value of the tokens in $U$ w.r.t.\ the end of epoch $j$ is
at most $n (i+1)/2^{2\log^{a}n}$.
\end{enumerate}
\end{enumerate}

For an epoch $0 \le j$ and a phase $0 \le i < \log^a n$ in this epoch, let
$p(j,i) = j\log^a n + i$ denote the global index of this phase.
We show that \whp
the condition $\PhaseInvA(j,i)$ holds at the beginning of each phase \mbox{$p(j,i) \le p_c$}.

\begin{sloppy}

\begin{lemma}\label{Lem:phaseInv}
For arbitrary $0 \le j$ and $0 \le i \le \log^a n -1$ such that $p(j,i) \le p_c$,
assume that the condition $\EpochInv(j)$ holds at some (global) time step $t$ and
the condition $\PhaseInvA(j,i)$ holds at the step $t_i = t + i(C/2) n \log^{1-a} n$.
Then the following conditions hold, where
\mbox{$t_{i+1} = t + (i+1)(C/2) n \log^{1-a} n$.}
\begin{enumerate}
\item
If $p(j,i) < p_c$, then \whp at step $t_{i+1}$
the condition $\PhaseInvA(j,i+1)$ holds.
\item
If $p(j,i) = p_c$, then \whp at step $t_{i+1}$
the total value, \wrt the end of epoch $j$,
of the minority-opinion tokens is $O(n\log n/2^{2\log^{a}n})$.
\end{enumerate}
\end{lemma}

\end{sloppy}

Lemma~\ref{Lem:phaseInv} is proven by analyzing the cancellations and duplications of tokens in one phase.
Lemma~\ref{Lem:epochinv} is proven by applying inductively Lemma~\ref{Lem:phaseInv}.
In turn, Theorem~\ref{Thm-FastExactMajorityA} below, which states
the $O(\log^{5/3} n)$ bound on the completion time of the \FastExactMajorityA protocol,
can be proven by applying inductively Lemma~\ref{Lem:epochinv}.


\begin{theorem}\label{Thm-FastExactMajorityA}
The \FastExactMajorityA protocol uses $\Theta(\log^2 n)$ states,
computes the majority \whp within $O(\log^{5/3} n)$ time
and reaches the correct all-$\done$ configuration or the all-$\fail$ configuration within
the expected
$O(\log^{5/3} n)$ time.
\end{theorem}

\begin{corollary}
The majority can be computed with $\Theta(\log^2 n)$ states
in $O(\log^{5/3} n)$ time \whp and in expectation.
\end{corollary}

\ignore{
In the proofs we use the following version of Chernoff Bound.

\begin{theorem} {\rm [Chernoff bound]}
Let $S_m=X_1+X_2+\dots+X_n$, where $X_i$, for $i = 1, 2, \ldots, n$, are independent random variables,
$0\leq X_i\leq 1$, $\E X_i=\mu_i$ and
$\mu=\mu_1+\mu_2+\dots +\mu_m$.
Then for $0\ge\varepsilon\le1$ and $\delta \ge 1$,
\begin{eqnarray*}
\Pr(|S_m - \mu| \ge \varepsilon\mu) & \leq & 2\exp \left\{-\frac{\varepsilon^2 \mu }{3} \right\}, \\
\Pr(S_m - \mu \ge \delta\mu) & \leq & \exp \left\{-\frac{\delta\mu}{3} \right\}.
\end{eqnarray*}
\end{theorem}
}

We give now some further explanations of the structure of our analysis,
referring the reader to
\ifthenelse{\isundefined{\ConferenceVersion}}{Appendix}{the full version~\cite{fullversion}}
for the formal proofs.
Lemma~\ref{lem_concentration_simple} and Claim~\ref{claim_close}
show the synchronization of the nodes which we rely on in our analysis.
Lemma~\ref{lem_concentration_simple} is used in the proof of Lemma~\ref{Lem:epochinv}, where
we analyze the progress of the computation through one epoch consisting of $O(n\log n)$ interactions
($O(\log n)$ parallel steps). Lemma~\ref{lem_concentration_simple} can be easily proven using first
Chernoff bounds for a single node and then the union bound over all nodes.
The proof of Claim~\ref{claim_close} is considerably more involved, but
we need this claim in the proof of Lemma~\ref{Lem:phaseInv}, where we look at the finer scale
of individual phases and have to consider intervals of $\Theta(\log^{1-a} n)$ interactions of a given node.
This claim shows, in essence, that most of the nodes stay tightly synchronized when they move from phase to phase
through one epoch. The $\epochstep$ counters of these nodes stay in a range of size
at most $c\log^{1-a} n$.

\begin{lemma}
\label{lem_concentration_simple}
For each sufficiently large constant $C$ and for $c = C^{3/4}$,
during a sequence of $t \le 2C n\log n$ interactions, $\mbox{\whp}(C)$\footnote{%
$\mbox{\whp}(\b)$ -- with probability at least $1 - n^{-\a(\b)}$,
where $\a(\b)$ grows to infinity with increasing $\b$.}
the number of interactions of each node is within $c\log n$ from the expectation of $2t/n$ interactions.
\end{lemma}

\begin{claim}
\label{claim_close}
For a fixed $j\ge 0$, assume that $\EpochInv(j)$ holds at
a time step $t$.
Let $W\subseteq V$ be the set of $n(1-o(1))$ nodes which satisfy at this step
the condition~\ref{EpochInvNormalNodes} of $\EpochInv(j)$
(that is, $W$ is the set of nodes which are in epoch $j$ with
$\epochstep$ counters at most  $c \log^a n$).
Then at an arbitrary but fixed time step $t < t' \le t + (3/4)C n\log n$, \whp\
all nodes in $W$ are in epoch $j$ and
all but $O(n/2^{6 \log^a n})$
of them have their $\epochstep$ counters within  $c/2 \cdot \log^{1-a} n$ from
$2(t'-t)/n$.
\end{claim}

Lemmas \ref{lem_broadcast_simple} and~\ref{lem_broadcast} describe the performance of the broadcast process in
the population-protocol model.
Lemma~\ref{lem_broadcast_simple} has been used before and is proven, for example,
in~\cite{DBLP:conf/podc/BilkeCER17}.
Lemma~\ref{lem_broadcast} is a more detailed view at the dynamics of the broadcast process,
which we need it in the context of Lemma~\ref{Lem:epochinv}
to show that the synchronization at the end epoch $j$ gives \whp $\EpochInv(j+1)$.

\begin{lemma}
\label{lem_broadcast_simple}
For each sufficiently large constant $c$,
the broadcast completes $\mbox{\whp}(c)$ within $cn\log n$ interactions.
\end{lemma}

\begin{lemma}
\label{lem_broadcast}
Let $b$ be any constant in $(0,1)$
and let $c_1$ and $c$ be sufficiently large constants.
Consider the broadcast process and let $t_1$
be the first step when $n/2^{6\log^b n}$  nodes are already informed
and $t_2 = t_1 + c_1 n \log^b n$.
Then the following conditions hold.
\begin{enumerate}
\item
With probability at least $1 - n^{-\omega(1)}$,
$n-O(n/2^{6\log^b n})$ nodes receive the message for the first time
within the $c_1 n \log^b n$ consecutive interactions
$\{ t_1+1, t_1+2, \dots , t_2 \}$.
\item
$\mbox{\Whp}(c)$, $t_1 \le c n \log n$
and no node interacts more than $4c \log n$ times within interval $[1, t_2]$.
\item
With probability at least $1 - n^{-\omega(1)}$,
there are  $n - O(n/2^{6\log^b n})$ nodes which interact within interval
$[t_1+1, t_2]$ at least $c_1\log^b n$ times but not more than $3 c_1\log^b n$ times.
\end{enumerate}
\end{lemma}


\section{Reducing the number of states to $\Theta(\log n)$}
\label{Sec-FastOptStates}

Our \FastExactMajorityA protocol described in Section~\ref{Sec:protocol}
requires $\Theta(\log^2 n)$ states per node.
Using the idea underlying the constructions of {\em
leaderless phase clocks} in~\cite{GP16} and~\cite{DBLP:conf/soda/AlistarhAG18}, we now
modify \FastExactMajorityA into the protocol $\FastExactMajorityB$, which still
works in $O(\log^{5/3} n)$ time but has only the asymptotically
optimal $\Theta(\log n)$ states per node.%
\footnote{Note that using the phase clock from~\cite{DBLP:conf/soda/GasieniecS18} would
not result in fewer states being needed for our protocol.}
The general idea is to separate from the whole population a subset of {\em clock nodes\/},
whose only functionality is to keep the time for the whole system.
The other nodes work on computing the desired output and check whether they
should progress to the next stage of the computation when they interact with clock nodes.
We note that while we use similar general structure and terminology as
in~\cite{DBLP:conf/soda/AlistarhAG18}, the meaning of some terms and the
dynamics of our phase clock are somewhat different. A~notable
difference is that in~\cite{DBLP:conf/soda/AlistarhAG18} the clock nodes keep
their time counters synchronized on the basis of the power of two choices in
load balancing: when two nodes meet, only the lower counter is incremented. In
contrast, we keep the updates of time counters as in the \ExactMajority and
\FastExactMajorityA protocols: both interacting clock nodes increment their
time counters, with the exception that the slower node is pulled up to the next
$\Theta(\log n)$-length phase or epoch, if the faster node is already there.

The nodes in the \FastExactMajorityB protocol are partitioned into two sets
with $\Theta(n)$ nodes in each set. One set consists of {\em worker nodes},
which may carry opinion tokens and work through canceling-doubling phases to
establish the majority opinion. These nodes maintain only information whether
they carry any token, and if so, then the value of the token (equivalently, the
age of the token, that is, the number of times this token has been split). Each
worker node has also a constant number of flags which indicate the current
activities of the node (for example, whether it is in the canceling stage of a
phase), but it does not maintain a detailed step counter. The other set
consists of {\em clock nodes}, which maintain their detailed epoch-step
counters, counting interactions with other clock nodes modulo $2C\log n$, for a
suitably large constant~$C$, and synchronizing with other clocks by the
broadcast mechanism at the end of epoch. Thus the clock nodes update their
counters in the same way as all nodes would update their counters in the
\FastExactMajorityA protocol%
\ifthenelse{\isundefined{\ConferenceVersion}}{%
, so Lemma~\ref{lem_broadcast} applies with some
obvious adaptation (the number of all nodes $n$ changes to the number of clock
nodes $n_c = \Theta(n)$ and only interactions between clock nodes are counted).
}{%
.}

The worker nodes interact with each other in a similar way as in
\FastExactMajorityA, but now to progress orderly through the
computation, they rely on the relatively tight synchronization of clock nodes.
A worker node $v$ advances to the next part of the current phase (or to the
next phase, or the next epoch), when it interacts with a clock node whose clock
indicates that $v$ should progress.
There is also the third type of nodes,
the {\em terminator nodes}, which will appear later in the computation.
A worker or clock node becomes a terminator node when it enters a $\done$ or $\fail$ state.
The meaning and function of these special states are as in protocols
\ExactMajority and \FastExactMajorityA.
\ifthenelse{\isundefined{\ConferenceVersion}}{%
In the Appendix we show  how to convert  a majority input instance
into the required initial workers-clocks configuration.
}{%
In the full version~\cite{fullversion} we show how to convert  a majority input instance
into the required initial workers-clocks configuration.
}

\begin{sloppy}

Referring to the state space of the \FastExactMajorityA protocol,
in the \FastExactMajorityB protocol
each worker node $v$ maintains data fields $v.\token$, $v.\epoch$ and $v.\ageinepoch$\
to carry information about tokens and their ages,
and a constant number of flags to keep track of the status of the node and its progress through
the current epoch and the current phase.
These include the status flags from the \FastExactMajorityA protocol
$v.\doubled$, $v.\outofsync$ and $v.\additionalepoch$, and flags indicating the progress:
the $v.\epochpart$ flag from \FastExactMajorityA and a new (multi-valued) flag
$\stage \in \{ \beginningphase, \canceling, \middlephase, \doubling, \phaseend \}$.
The clock nodes maintain the $\epochstep$ counters.
The nodes have constant number of further flags, for example to support the initialization to workers and clocks and
the implementation of the additional epoch and the slow backup protocol.
Thus in total each node has only $\Theta(\log n)$ states.

\ifthenelse{\isundefined{\ConferenceVersion}}{%
Further details of \FastExactMajorityB, including pseudocodes and outline of
the proof of Theorem~\ref{thm-xxxx}
which summarizes the performance of this protocol, are given in the Appendix.
}{%
Further details of \FastExactMajorityB, including pseudocodes and outline of
the proof of Theorem~\ref{thm-xxxx}
which summarizes the performance of this protocol, are given in the full version~\cite{fullversion}.
}
\end{sloppy}

\begin{theorem}\label{thm-xxxx}
The \FastExactMajorityB protocol uses $\Theta(\log n)$ states,
computes the exact majority \whp within $O(\log^{5/3} n)$ parallel time
and stabilizes (in the correct all-$\done$ configuration or in the all-$\fail$ configuration) within
the expected
$O(\log^{5/3} n)$ parallel time.
\end{theorem}

\begin{corollary}
The exact majority can be computed with $\Theta(\log n)$ states
in $O(\log^{5/3} n)$ parallel time \whp and in expectation.
\end{corollary}


\bibliography{refs}

\ifthenelse{\isundefined{\ConferenceVersion}}{}{\end{document}\endinput}

\newpage

\appendix
\section{Appendix}

\subsection{Pseudocodes for Section \ref{generalidea} -- $O(\log^2 n)$-time  \ExactMajority protocol}

This section contains our pseudocodes left out from Section \ref{generalidea}
for our \ExactMajority protocol.

\RestyleAlgo{boxruled}
\LinesNumbered
\SetEndCharOfAlgoLine{}

\begin{algorithm}[h]
\uIf{$v.\normal \wedge u.\normal$}{
\lIf{$\neg \consistent(v.\Time, u.\Time)$}{
    $v.\fail, u.\fail \Let \true$;
}
\uElse{
\tcp{$\consistent(v.\Time, u.\Time) \Rightarrow |v.\phase - u.\phase| \le 1$}
\uIf{$v.\phase \neq u.\phase$}{
    let $v$ be the node in the higher phase, that is, $v.\phase = u.\phase +1$; \;
    \tcp{$u$ in the final part of a phase and $v$ in the next phase}
    $u.\phasestep \Let (2C\log n) - 1$; \tcp*{$u $ moves to next phase at $\nextstep(u)$}
}
\;
\uElseIf{$v.\phase = u.\phase$}{
\uIf{both $v$ and $u$ in the canceling stage and have opposite tokens}{
            $u.\token \Let \notoken$;~ $v.\token \Let \notoken$\;
}
\uElseIf{$v$ and $u$ in the doubling stage, exactly one $x\in \{v,u\}$ has a token,
     and that token hasn't doubled yet ($\neg x.\doubled$)}{
        let $v$ be the node with a token;\;
        $u.\token \Let v.\token$;~
        $v.\doubled, u.\doubled \Let \true$;
}

}
\;
\lFor {each $x \in \{ u,v\}$}{
        \{ $x.\Time \Let x.\Time + 1$;~ $\nextstep(x)$; \}
}
}
}
 \caption{\ExactMajority protocol -- interaction of two nodes $v$ and $u$\label{BasicAlgo}}
\end{algorithm}

\begin{algorithm}[h]
$\nextstep(x)$:\;
\bindent
   \lIf{$x.\phase = \log n + 2$}{$x.\fail\Let \true$; }
   \uElseIf{$(x.\token \neq \notoken \wedge x.\phasestep = 0)$}{
                  \tcp{the token at $x$ has completed a phase}
                  \uIf{$x.\doubled$}{
                            \tcp{everything's OK: the token progresses normally}
                            $x.\doubled \Let \false$;
                  }
                  \uElse{
                            \tcp{the token has failed to split in the last phase}
                            $x.\done \Let \true$;\;
                  }
              }
\eindent
 \caption{\ExactMajority protocol - end of interaction, node $x$ progresses to the next step\label{BasicNextStepCode}}
\end{algorithm}

\newpage
\subsection{Pseudocodes for Section \ref{Sec:protocol} -- protocol \FastExactMajorityA}
This section contains our pseudocodes left out from Section \ref{Sec:protocol}
for our \FastExactMajorityA protocol.

\begin{algorithm}[!h]
\uIf{$v.\normal \wedge u.\normal$}{
\lIf{$\neg \consistent(v.\Time, u.\Time)$}{
    $v.\fail, u.\fail \Let \true$;
}
\uElse{
\;
\uIf{$v.\epoch \neq u.\epoch$}{
    let $v$ be the node in the higher epoch, that is, $v.\epoch = u.\epoch +1$; \;
    \tcp{$u$ in the final part of an epoch and $v$ in the next epoch}
    $u.\epochstep \Let (2C\log n) - 1$;\;
}
\;
\uElseIf{$((v.\epochpart = u.\epochpart = 0) \wedge (v.\phase = u.\phase))$}{
\uIf{both $v$ and $u$ in the canceling stage and have opposite tokens}{
            $u.\token \Let \notoken$;~ $v.\token \Let \notoken$\;
}
\uElseIf{both $v$ and $u$ in the doubling stage and exactly one has a token}{
        let $v$ be the node with a token (the other case is symmetric);\;
        \uIf{$\neg v.\doubled$}{
            $u.\token \Let v.\token$;\;
            $u.\ageinepoch \Let v.\ageinepoch$; \tcp*{$ = v.\phase = u.\phase$}
            $v.\doubled, u.\doubled \Let \true$;
        }
}
}
\;
\lFor {each $x \in \{ u,v\}$}{
     \{    $x.\Time \Let x.\Time + 1$;~ $\nextstep\_\mbox{\it normal}(x)$; \}
}
}
}
 \caption{\FastExactMajorityA protocol -- interaction of two normal nodes $v$ and $u$\label{AlgoNormalProgress}}
\end{algorithm}

\begin{algorithm}[!h]
\uIf{$v.\outofsync \vee u.\outofsync$}{
   \lIf{$\neg \consistent(v.\Time, u.\Time)$}{
       $v.\fail, u.\fail \Let \true$;
   }
   \uElse{
       \;
       \uIf{$v.\epoch \neq u.\epoch$}{
           let $v$ be the node in the higher epoch, that is, $v.\epoch = u.\epoch +1$; \;
           \tcp{$u$ in the final part of an epoch and $v$ in the next epoch}
          $u.\epochstep \Let (2C\log n) - 1$;\;
          \;
      }
     \uElseIf{$\neg(v.\outofsync \wedge u.\outofsync)$}{
         let $v$ be the node with an out-of-sync token; \;
         \uIf{$((v.\ageinepoch < \log^{a}n) \wedge (u.\token = \notoken))$}{
            \tcp{split the $v$'s token}
             $u.\token \Let v.\token$;~ $u.\outofsync \Let \true$;\;
             $v.\ageinepoch \Let v.\ageinepoch + 1$;~  $u.\ageinepoch \Let v.\ageinepoch$;\;
             \uIf{$v.\ageinepoch = \log^{a}n$}{
                  \lFor{each $x \in \{ u,v\}$}{
                      \lIf{$x.\epochpart = 1$}{$x.\outofsync \Let \false$;}
                  }
             }
        }
      }
     \For{each $x \in \{ u,v\}$}{
        $x.\Time \Let x.\Time + 1$;\;
        \lIf{$x.\outofsync$}{
            $\nextstep\_\mbox{\it outofsync}(x)$;~ {\bf else}~ $\nextstep\_\mbox{\it normal}(x)$;
        }
     }
   }
}
 \caption{\FastExactMajorityA protocol -- interaction of an out-of-sync node \label{AlgoOutofsync}}
\end{algorithm}

\newpage

\begin{algorithm}[h]
$\nextstep\_\mbox{\it normal}(x)$:\;
\bindent
   \lIf{$x.\epoch = \log^{1-a}n +2$}{$x.\fail\Let \true$; }
   \uElseIf{$(x.\token \neq \notoken \wedge x.\phasestep = 0)$}{
             \uIf{$(x.\epochpart = 0 \wedge x.phase > 0) \vee (x.\epochpart = 1 \wedge x.phase = 0)$}{
                  \tcp{a normal token has completed a phase}
                  \uIf{$x.\doubled$}{
                            \tcp{everything's OK: the token progresses normally}
                            $x.\doubled \Let \false$; ~ $x.\ageinepoch \Let  x.\ageinepoch + 1$;\;
                  }
                  \uElse{
                            \tcp{the token has failed to split in the last phase}
                            $x.\outofsync \Let \true$;\;
                  }
              }
         \uElseIf{$x.\epochstep = 0$}{
              \tcp{a normal token has completed an epoch}
              $x.\ageinepoch \Let 0$;
         }
                   }
\eindent
 \caption{\FastExactMajorityA protocol -- End of interaction of normal node $x$\label{NextStepCodeNormal}}
\end{algorithm}

\begin{algorithm}[h]
$\nextstep\_\mbox{\it outofsync}(x)$:\;
\tcp{node $x$ has an out-of-sync token}
\bindent
   \lIf{$x.\epoch = \log^{1-a}n + 2$}{$x.\fail\Let \true$; }
   \uElseIf{$((x.\ageinepoch = \log^{a}n) \wedge (x.\epochpart = 1))$}{
              \tcp{out-of-sync token reached the final age
              and entered the second part of epoch}
              $x.\outofsync \Let \false$;\;
        }
        \uElseIf{$(x.\epochstep = 0)$}{
                  \tcp{out-of-sync token has completed an epoch (so it has failed to split)}
                   $x.\additionalepoch \Let \true$;\;
        }
\eindent
 \caption{\FastExactMajorityA protocol -- end of interaction of out-of-sync node $x$ \label{NextStepCodeOutofsync}}
\end{algorithm}

\newpage
\subsection{Proofs for Section \ref{analysis} -- protocol \FastExactMajorityA}

For convenience we assume in the proofs that opinion $A$ is the majority opinion, that is,
$a_0 > b_0$.

\begin{proofof}{Lemma \ref{Lem:epochinv}
}
We consider an epoch $j \ge 0$ such
that phase $p_c$ belongs to this or a later epoch
and assume that the condition $\EpochInv(j)$
holds at a (global) step $t$.

\paragraph{{\it{Case 1: phase $p_c$ belongs to a later epoch $j' > j$.}}} $\;$

\noindent
Applying Lemma~\ref{Lem:phaseInv} inductively to phases $i = 0, 1, \ldots, \log^a n - 1$
of epoch $j$, we conclude that \whp the condition $\PhaseInvA(j,\log^a n)$ holds
at step $t' = t + (C/2) n \log n$. (Using the definition of step $t_i$ from Lemma~\ref{Lem:phaseInv},
$t' = t_i$, for $i = \log^a n$.)

Assume therefore that the condition $\PhaseInvA(j,\log^a n)$ holds
at step $t'$.
Thus for each node $v\in V$, $v$ is in epoch $j$ and
$|v.\epochstep - C\log n| \le 4c\log n$.
Actually, for most of the nodes $v$, $0 \le  v.\epochstep - C\log n \le  c\log^{1-a} n$
(from the condition~\ref{hjjx90sa} of $\PhaseInvA(j,\log^a n)$), but there may be
a small number of nodes with their $\epochstep$ counters outside this range.

At step $t'$, the total value, \wrt the end of epoch $j$,
of the tokens which are out-of-sync or in nodes with
$\epochstep$ counters outside the interval $C\log n + [0, c\log^{1-a} n]$
is at most $n \log n /2^{2\log^{a}n}$
(from the condition~\ref{hk892a} of $\PhaseInvA(j,\log^a n)$).
We first wait until the step $t'' = t' + (5/2)cn\log n$ to ensure that \whp
all nodes are in the second part of the epoch.
Indeed, the conditions of the system at step $t'$ and
Lemma~\ref{lem_concentration_simple} applied to steps $t, t+1, \ldots, t'$ give that
at step $t''$ \whp all $\epochstep$ counters are within
the interval $C\log n + [0, 10c\log n]$.

At step $t''$ most of the tokens are normal, that is, their value is $1/2^{(j+1)\log^a n}$ as required
at the end of epoch $j$ (and at the beginning of the next epoch $j+1$).
The out-of-sync tokens have values larger than $1/2^{(j+1)\log^a n}$,
but at most $1/2^{j\log^a n}$,
and their total value is at most $n \log n /2^{2\log^{a}n}$
\wrt the end of epoch $j$.
We view the set of out-of-sync tokens as the set $\cT$ of {\em base tokens\/}
of value $1/2^{(j+1)\log^a n}$,
which are grouped into larger tokens.
That is, an out-of-sync token of value $1/2^{(j+1)\log^a n - i}$, for some $1 \le i \le \log^a n$,
is a group of $2^i$ base tokens.
The number of base tokens is at most $n \log n /2^{2\log^{a}n}$.
We consider an arbitrary base token $b$ and show
that \whp this token interacts in steps $t'', t''+1, \ldots, t''' = t+((3/4)C + 3c)n\log n$
with at least $\log^a n$ empty nodes.
This will imply that by the step $t'''$ \whp
the base token $b$ completely splits off from its initial larger token
and becomes a separate individual
token of value $1/2^{(j+1)\log^a n}$ (as required at the end of epoch $j$).
(If the base token $b$ is initially part of a token of value $1/2^{(j\log^a n) + i}$,
for some $0 \le i \le \log^a n - 1$, then it becomes a separate token after $\log^a n - i$
interactions with empty nodes.)
By the union bound, by the step $t'''$ \whp there are no out-of-sync tokens left, that is,
all tokens are normal tokens of value $1/2^{(j+1)\log^a n}$.

Token $b$ interacts \whp with at least $\log^a n$ empty nodes during the interval
$[t'', t''']$ because $t''' - t'' \ge (C/4)n\log n$ and
\whp there are $n/10$ empty nodes at each step in this interval.
This follows from Lemma~\ref{cancellations} applied to the last phase of the epoch
(phase $\log^a n -1$).
This lemma implies that
\whp the cancellation stage of this phase reduces the number of normal tokens of value
$1/2^{(j+1)(\log^a n) \, - 1}$ to at most $(4/10)n$, while the total value,
\wrt to the end of the epoch, of the other tokens is $o(n)$.
This means that \whp during the interval
$[t'', t''']$ the number of tokens stays \whp below $(8/10 + o(1)) n$, so
the number of empty nodes is at least $n/10$.
Thus at each step of this interval the probability that token~$b$ interacts with an empty node
or there are fewer than $n/10$ empty nodes is at least $2\cdot (1/n) \cdot (1/10) = 5/n$.
There are $t''' - t'' = (C/4 + c/2) n\log n$ steps in this interval,
so the Chernoff bound~\eqref{ChernoffC}
implies that for sufficiently large constant $C$,
token $b$ interacts \whp with at least $\log n$ empty nodes.

Summarizing what we have established so far,
at step $t''' = t+((3/4)C + 3c) n\log n$ \whp all tokens are normal tokens of value $1/2^{(j+1)\log^a n}$.
Moreover, since  at step $t''$ \whp all nodes are in epoch $j$ and their $\epochstep$ counters are within
the interval $C\log n + [0, 10c\log n]$, then at step $t'''$ \whp
all nodes are in epoch $j$ and their $\epochstep$ counters are within
the interval $(3/2)C\log n + [0, 12c\log n]$ (using Lemma~\ref{lem_concentration_simple}).
This implies that \whp
at some step $t^{(4)} \le t''' + (1/4)Cn\log n$,
the first node enters the next epoch $j+1$ and initiates
the broadcast that pulls up all nodes to epoch $j+1$.
We now use Lemma~\ref{lem_broadcast} with $b = a$ and $c_1 = c/3$ to conclude
that \whp
the condition $\EpochInv(j+1)$ holds
at some step $\tilde{t} \le t^{(4)} + 2cn\log n \le t+(C + 5c)n\log n$.
(This step $\tilde{t}$ is $t_2$ steps from the beginning of the broadcast, where $t_2$ is defined
in Lemma~\ref{lem_broadcast}.)

\paragraph{{\it{Case 2: both phases $p_c$ and $p_c+1$ belong to epoch $j$.}}} $\;$

\noindent
Let phase $p_c$ be phase $q$ of epoch $j$, for some $0 \le q \le \log^a n - 2$.
Applying Lemma~\ref{Lem:phaseInv} inductively to phases $i = 0, 1, \ldots, q$,
we conclude that at the step $t_{q+1} = t + (q+1)(C/2) n \log^{1-a} n \le t+(C/2)n\log n$
the total value, \wrt the end of epoch $j$,
of the minority-opinion tokens is \whp $O(n\log n/2^{2\log^{a}n}) = o(n)$.
Thus the total value, \wrt the end of epoch $j$, of the tokens which can cancel out
after the step $t_{q+1}$ is only $o(n)$.
We also know (from the condition $\EpochInv(j)$ at step $t$
and using Lemma~\ref{lem_concentration_simple}) that \whp at the step $t_{q+1}$
each node is in epoch $j$ or $j-1$.

At step $t_{q+1}$,
the total value, \wrt (the beginning of) phase $q$ of epoch $j$,
of the majority-opinion tokens is at least $n/3$
(from the definition of phase $p_c$), so this total value is at least $(4/3)n$
\wrt to the end of phase $q+1$ and hence also at least $(4/3)n$ \wrt to the end of epoch $j$.
Since \whp only $o(n)$ of this total can cancel out after step $t_{q+1}$,
\whp from step $t_{q+1}$ on the total
value of the majority-opinion tokens remains at least $(4/3 - o(1))n$
(\wrt the end of epoch $j$).
Thus \whp not all tokens split by the end of epoch $j$
to the value required at the end of this epoch, or
otherwise we would have tokens in epoch $j+1$ of the total value
(\wrt the beginning of epoch $j+1$) at least $(4/3 - o(1))n$.
This means that there must be out-of-sync tokens which reach the end of epoch $j$ and enter
the $\additionalepoch$ state.
It remains to provide a bound on the step when this happen for the first time.

\Whp
at the step $t+(3/4)C n \log n$
all nodes are in the second part of epoch $j$ (from the condition $\EpochInv(j)$ at step $t$
and using Lemma~\ref{lem_concentration_simple}) and
by the step $t + (C +c +o(1)) n \log n$ all nodes reach the end of epoch $j$.
Indeed, since at step $t$ all but $o(n)$ nodes are in epoch $j$,
while those remaining $o(n)$ nodes are in epoch $j-1$,
the first node reaches the end of epoch $j$
within $(C+o(1))n \log n$ steps and then at most $c n \log n$ further
steps would take all nodes to the end of epoch $j$ (Lemma~\ref{lem_broadcast_simple}).
Thus for the step $\tilde{t}$ when the first out-of-sync token reaches the end of epoch $j$,
\whp $t+(3/4)C n \log n \le \tilde{t} \le t + (C+c + o(1))n\log n$.

\paragraph{{\it{Case 3: phases $p_c$ is the last phase of epoch $j$.}}} $\;$

\noindent
Phase $p_c$ is the last phase $q = \log^a n - 1$ in epoch $j$.
Similarly to Case 2,
we apply Lemma~\ref{Lem:phaseInv} inductively to phases $i = 0, 1, \ldots, q$ to
conclude that at the step $t_{q+1} = t + (C/2) n \log n $
the total value, \wrt the end of epoch $j$,
of the minority-opinion tokens is \whp $O(n\log n/2^{2\log^{a}n}) = o(n)$.
The total value of the majority-opinion tokens, \wrt the end of epoch $j$ or, equivalently,
the end of phase $p_c$,
is at least $(2/3)n$ (from the definition of phase $p_c$).
If the difference between the total value of the majority-opinion tokens and the minority-opinion tokens is greater than $n$,
then, similarly to Case 2, there will have to be an out-of-sync token reaching the end of epoch $j$
and entering the
$\additionalepoch$ state.
This may also happen
if this difference is less than $n$ but too close to $n$, so eventually too few empty nodes
to allow sufficient chance for all tokens
to split to the value required at the end of epoch $j$.
If there are out-of-sync tokens reaching the end of epoch $j$, then, as in Case 2,
\whp the first out-of-sync token reaches the end of epoch $j$ at a step
$\tilde{t} \le t + (C+c + o(1))n\log n$.

If all tokens do split by the end of epoch $j$ to the required value,
then all nodes progress to the next epoch $j+1$.
The difference between the total value of the majority-opinion tokens and the minority-opinion tokens
\wrt to the end of epoch $j+1$ is at least $(2/3)n \cdot 2^{\log^a n} > n$,
so there must be an out-of-sync token which reaches the end of epoch $j+1$
and enters the
$\additionalepoch$ state.
\Whp all nodes reach the end of this epoch by the step
$t + (2C +2c +o(1)) n \log n$.
Thus \whp there is a step
$\hat{t} \le t + (2C+2c + o(1))n\log n$
when the first out-of-sync node reaches the end of epoch $j+1$ and enters the $\additionalepoch$ state.
\end{proofof}

\begin{proofof}{Lemma \ref{Lem:phaseInv}}
We consider first the case when $p(j,i) < p_c$.
That is, we consider a phase $i$ of epoch $j$
such that
$2^{p(j,i)} |a_0-b_0| \leq n/3$,
assume that $\EpochInv(j)$ holds at a step $t$ and
$\PhaseInvA(j,i)$ holds at step $t_i = t + i(C/2) n \log^{1-a} n$, and
show that \whp $\PhaseInvA(j,i+1)$ holds at step
$t_{i+1} = t_{i} + (C/2) n \log^{1-a} n$.

The assumptions of the lemma imply
that at step $t_i$ all tokens have values at most $1/2^{j\log^{a} n}$
and at least $1/2^{(j+1)\log^{a} n}$.
Furthermore,
the set $W_i$ of nodes which are normal and in
the beginning part of phase $i$ (in epoch $j$) has size
at least $n(1 - (i+1)/2^{2\log^{a} n})$.
(Set $W_i$ is the set $W$ defined in $\PhaseInvA(j,i)$ at step $t_i$.)
All tokens in $W_i$ have value $1/2^{j(\log^{a} n) + i} = 1/2^{p(j,i)}$.
The total value, \wrt the end of epoch $j$, of tokens in $V \setminus W_i$ is
at most $n(i+1)/2^{2\log^{a} n})$.

In Lemma~\ref{cancellations} we analyze cancellations of tokens
in phase $i$ by step $t'_i = t_i + (C/4 - c/2)n\log^{1-a} n$ and
in Lemma~\ref{Lem:phaseduplications} we analyze doubling of tokens in this phase by step
$t''_i =  t_i + (C/2 - c/2)n\log^{1-a} n < t_{i+1}$.
More precisely, in Lemma~\ref{cancellations} we analyze cancellations of tokens
in steps $t_i, t_i + 1, \ldots,  t'_i$, when \whp most of the
nodes stay in the cancellation stage of phase $i$.
In Lemma~\ref{Lem:phaseduplications} we analyze doubling of tokens in steps
$t'''_i = t_i + (C/4 + c/2)n\log^{1-a} n, t'''_i + 1, \ldots, t''_i$, when \whp most of the
nodes stay in the doubling stage of phase $i$.

Lemma~\ref{cancellations} says that \whp
the token cancellations in steps $t_i, t_i + 1, \ldots,  t'_i$
result in at least $(6/10)n$ empty nodes at step $t'_i$.
Moreover, at step $t'_i$, the set $W'$ of nodes which are normal and in
the end part of the cancellation stage of phase $i$ has size \whp
at least $n(1 - (i+1+o(1))/2^{2\log^{a} n})$, and
the total value, \wrt the end of epoch $j$, of tokens in $U' = V \setminus W'$ is \whp
at most $n(i+1+o(1))/2^{2\log^{a} n})$.

Lemma~\ref{Lem:phaseduplications} assumes the state of the system which Lemma~\ref{cancellations}
guarantees \whp, and shows that \whp by step $t''_i$ most of the tokens have split to the value
$1/2^{p(j,i)+1}$.
More precisely, at step $t''_i$,
the set $W''$ of
nodes which are normal, in the end part of phase $i$ and either empty
or with tokens of value $1/2^{p(j,i)+1}$
has size \whp at least $n(1 - (i+1+o(1))/2^{2\log^{a} n})$.
The total value, \wrt the end of epoch $j$, of the tokens in $U'' = V \setminus W''$ is \whp
at most $n(i+1+o(1))/2^{2\log^{a} n})$.

Let $\tilde{W} \subseteq W''$ be the set of nodes in $W''$ which
at step $t_{i+1}$ are normal and in the beginning part of phase $i+1$.
Claim~\ref{claim_close} implies that \whp
$|\tilde{W}| \ge |W''| - O(n/2^{6 \log^a n}) \ge n(1 - (i+2)/2^{2\log^{a} n})$.
Let $W_{i+1}$ be the set defined in the condition~\ref{hjjx90sa} of $\PhaseInvA(j,i+1)$ at step $t_{i+1}$.
Observing that $\tilde{W} \subseteq W_{i+1}$, we conclude that
\whp the condition~\ref{hjjx90sa} of $\PhaseInvA(j,i+1)$ holds at step $t_{i+1}$.

We consider now the tokens which are at step $t_{i+1}$ in
$\tilde{U} = V \setminus \tilde{W} = U'' \cup (W'' \setminus \tilde{W})$.
These tokens originate from the tokens which were at step $t''_i$ in $U''$ or in
$W'' \setminus \tilde{W}$.
The total value, \wrt the end of epoch $j$, of
these tokens is therefore \whp at most $n(i+1+o(1))/2^{2\log^{a} n})$
(the contribution from the tokens in $U''$) plus  $2^{\log^{a} n - i} \cdot O(n/2^{6 \log^a n})$
(the contribution from the tokens in $W'' \setminus \tilde{W}$).
Thus the total value, \wrt the end of epoch $j$, of the tokens which are at step $t_{i+1}$ in
$\tilde{U}$ is \whp at most $n(i+2)/2^{2\log^{a} n})$.
Observing that $V \setminus W_{i+1} \subseteq \tilde{U}$, we conclude that
\whp the condition~\ref{hk892a} of $\PhaseInvA(j,i+1)$ holds at step $t_{i+1}$.

Now we show that at step $t_{i+1}$ \whp the condition~\ref{k892a} of $\PhaseInvA(j,i+1)$ holds  as well.
Let $u\in V$ be an arbitrary node.
\Whp the number of interactions of node $u$
in steps $t, t+1, \ldots, t_{i+1}$
differs from $2(t_{i+1} - t)/n = (i+1)C\log^{1-a} n$ by at most  $c\log n$
(from Lemma~\ref{lem_concentration_simple}).
At step $t$, node $u$ was either within its first $3c\log n$ steps of epoch $j$
or was a normal node in the final quarter of the previous epoch $j-1$.
If the former, then \whp at step $t_{i+1}$, $u.\epochstep$ differs from $(i+1)C\log^{1-a} n$
by at most $4c\log n$.
If the latter, that is, if
$u$ was at step $t$ a normal node in the final quarter of epoch $j-1$,
then we consider two cases.

If $t_{i+1} \ge t + cn\log n$ (equivalently, $i+1 \ge (c/C)\log^a n$), then \whp node $u$ enters
epoch $j$ at some step $\t$, where $t \le \t \le t + cn\log n$ (from Lemma~\ref{lem_broadcast_simple}),
and
the number of $u$'s interactions in steps $\t+1, \t+2, \ldots, t_{i+1}$
is within $c\log n$ from  $2(t_{i+1} - \t)/n$ (from Lemma~\ref{lem_concentration_simple}).
Thus at step $t_{i+1}$, $u.\epochstep$ differs from
$2(t_{i+1} - \t)/n$ by at most $c\log n$, so it differs from
$(i+1)C\log^{1-a} n = 2(t_{i+1} - t)/n$
by at most $2(\t - t)/n + c\log n \le  3c\log n$.
If $t_{i+1} < t + cn\log n$ (that is, if $i+1 < (c/C)\log^a n$),
then if node $u$ enters epoch $j$ by step $t_{i+1}$, then
it has \whp at most $2(t_{i+1} -  t)/n + c\log n < 3c\log n$ interactions in this epoch, so
$u.\epochstep$ differs from $(i+1)C\log^{1-a} n = 2(t_{i+1} - t)/n$ by at most $3c\log n$.

Thus in all cases, \whp at step $t_{i+1}$ either node $u$ is still a normal node in the final quarter of
epoch $j-1$ and $i+1 < (c/C)\log^a n$,
or $u$ is in epoch $j$ and $|u.\epochstep - (i+1)C\log^{1-a} n| \le 4c\log n$.
By the union bound, \whp the condition~\ref{k892a} of $\PhaseInvA(j,i+1)$ holds at step $t_{i+1}$.
(We note that the bounds on the $\epochstep$ counters
given in this condition are satisfied by all nodes since the bounds in the condition~\ref{hjjx90sa}
are tighter.)

The second case of the lemma, that is, when phase $p(j,i)$ is phase  $p_c$,
is covered by Lemma~\ref{cancellations_small}.
We analyze in this lemma
token cancellations in phase $p_c$ and show that
at step $t'_i = t_i + (C/4 - c/2)n\log^{1-a} n < t_{i+1}$,
\whp the total value, \wrt the end of epoch $j$, of the minority-opinion tokens
is $O(n\log n/2^{\log^{a}n})$.
\end{proofof}

\begin{proofof}{Claim \ref{claim_close}}
At each interaction two nodes are
chosen uniformly at random.
We consider the sequence of $t'-t$ interactions at steps $t, t+1, \ldots t'-1$, which
can be modeled by the balls-into-bins process  placing randomly
$2(t'-t)$ balls
in $n$ bins (with bins corresponding to nodes).
There is a restriction that no two consecutive odd-even balls are placed in the same bin, since
each interaction is between two distinct nodes.
In this balls-into-bins process,
the occupancy $Z_k$ of bin $k$ has the expected value $\E(Z_k) = 2(t'-t)/n$
and it can be shown using Chernoff bounds that for any constant $\delta > 0$,
\begin{equation}\label{jkw2}
 \Pr(|Z_k - 2(t'-t)/n| \ge \delta \log^{1-a} n)
    \leq \exp\{- (\delta^2/(6C)) \cdot \log^{a} n\}.
\end{equation}
Indeed, $Z_k = Z_k^{(1)} + \cdots + Z_k^{(t'-t)}$,
where $Z_k^{(\t)}$ indicates whether any of the $\t$-th pair of balls is placed in bin $k$,
and $\E(Z_k^{(\t)}) = 2/n$.
Denoting $\mu = 2(t'-t)/n$, if $\delta \log^{1-a} n \le \mu$, then
applying~\eqref{ChernoffA} and noting that $\mu \le (3/2)C\log n$ and $a = 1/3$, we have
\begin{eqnarray*}
\Pr(|Z_k - \mu| \ge \delta \log^{1-a} n)
 &\leq& \exp\left\{-\frac{\mu}{4}\cdot\left(\frac{\delta\log^{1-a} n}{\mu}\right)^2 \right\}
 \; \leq \; \exp\left\{-\frac{1}{4}\cdot
      \frac{\delta^2\log^{2-2a} n}{(3/2)C\log n} \right\} \\
 &=& \exp\{- (\delta^2/(6C)) \cdot \log^{a} n\}.
\end{eqnarray*}
In the case when $\delta \log^{1-a} n > \mu$, we apply~\eqref{ChernoffB}
and get the bound $\exp\{- (\delta/3)\log^{1-a} n\}$, which is smaller than the bound~\eqref{jkw2}.
%

Inequality~\eqref{jkw2} with a sufficiently large $\d$
implies that the expectation of the number $X$ of nodes with interaction counts within  $\delta \log^{1-a} n$
from $2(t'-t)/n$
(equivalently, the expectation of the number $X$ of bins with such occupancies)
is at least $n(1 - 1/2^{6\log^{a} n})$.
We show that $X$ is concentrated around the expectation $\E(X)$,
but the argument
is not completely straightforward because the counts of interactions are not independent.
To deal with these dependencies, we modify the balls-to-bins process so that any bin which reaches
the load of $\sqrt[4]{n}$ balls is not considered in any subsequent selections of bins.

Let $\tilde{Z}_k$ denote the occupancy of bin $k$ in the modified process
and let
$\cE$ denote
the event that in the original process no bin receives more than $\sqrt[4]{n}$ balls.
By coupling the modified process with the original process for as long as no load of a bin exceeds $\sqrt[4]{n}$,
we have that $Z_k \neq \tilde{Z}_k$ implies $\overline{\cE}$.
Thus using~\eqref{jkw2} we obtain
\begin{eqnarray}
 \Pr(|\tilde{Z}_k - 2(t'-t)/n| \ge \delta \log^{1-a} n)
   & \le & \Pr(|{Z}_k - 2(t'-t)/n| \ge \delta \log^{1-a} n\; \mbox{or}\; \overline{\cE}) \nonumber \\
   & \le &  \Pr(|{Z}_k - 2(t'-t)/n| \ge \delta \log^{1-a} n) + \Pr(\overline{\cE}) \nonumber \\
   &  \leq & \exp\{- (\delta^2/(6C)) \cdot \log^{a} n\} + \exp\{ - \Theta(\sqrt[4]{n}) \} \nonumber \\
   & \leq & 2 ^{-6\log^a n}, \label{hjw7aq}
\end{eqnarray}
where the last inequality holds for sufficiently large $\d$.
Therefore the expectation of the number $\tilde{X}$ of bins in
 the modified process with load
within  $\delta \log^{1-a} n$
from $2(t'-t)/n$
is at least $n - n/2^{6\log^{a} n}$.

\begin{sloppy}

We consider random variables
$\tilde{X}_k = \E(\tilde{X}~|~\tilde{Z}_1, \dots , \tilde{Z}_k)$, for $k = 0,1,\ldots, n$.
The sequence $(\tilde{X}_0,\tilde{X}_1, \ldots, \tilde{X}_n)$ is a Doob martingale
with $\tilde{X}_0 = \E(\tilde{X})$ and $\tilde{X}_n = \tilde{X}$
(see, for example, \cite{Mitzenmacher:2005:PCR:1076315}).
For each $0 \le k \le n-1$ and any $z', z'' \in \{0,1,\ldots, \sqrt[4]{n}\}$, we have
\begin{equation}\label{bckj01l}
 |\E(\tilde{X}~|~\tilde{Z}_1, \dots , \tilde{Z}_{k}, \tilde{Z}_{k+1} = z')
        - \E(\tilde{X}~|~\tilde{Z}_1, \dots , \tilde{Z}_{k}, \tilde{Z}_{k+1} = z'')| \; \le \; \sqrt[4]{n} + 1,
\end{equation}
provided that the conditions in both expectations are feasible (that is, have positive probability).
To see this, consider the following
coupling of the generation of
the two conditional variables above,
that is the variable $\tilde{X}$ under the condition that $\tilde{Z}_1, \dots , \tilde{Z}_{k}, \tilde{Z}_{k+1} = z'$
and the variable $\tilde{X}$ under the condition that $\tilde{Z}_1, \dots , \tilde{Z}_{k}, \tilde{Z}_{k+1} = z''$.
Assuming $z' > z''$,
first put $Z_i$ balls in bin $i$ for each $i = 1, 2, \ldots, k$ and $z''$ balls in bin $k+1$.
Then split the remaining $y$ balls into two groups of $y - (z' - z'')$ balls and $z' - z''$ balls,
distribute the balls from the first group randomly in bins $k+2, k+3, \ldots n$, observing the $\sqrt[4]{n}$ bound on the load
of each bin, and {\em assign\/} (without putting in) the
balls from the second group randomly to bins $k+2, k+3, \ldots n$, again observing the upper bound on the load of each bin.
To get the value of the first variable, put the balls from the second group to the bins they were assigned to.
To get the value of the second variable,
put all balls from the second group to bin $k+1$.
The values of the two generated variables differ by at most $z' - z'' + 1 \le \sqrt[4]{n} + 1$.

\end{sloppy}

Inequality~\eqref{bckj01l} implies that
the expectation
$\E_{\tilde{Z}_{k+1}}(\E(\tilde{X}~|~\tilde{Z}_1, \dots , \tilde{Z}_{k}, \tilde{Z}_{k+1})) \equiv \tilde{X}_k$
does not differ from
$\E(\tilde{X}~|~\tilde{Z}_1, \dots , \tilde{Z}_{k}, \tilde{Z}_{k+1}) \equiv \tilde{X}_{k+1}$
by more than $\sqrt[4]{n} + 1$.
Thus
for each $0 \le k \le n-1$, $|\tilde{X}_{k+1} - \tilde{X}_{k}| \le \sqrt[4]{n} + 1$, so
the Azuma-Hoeffding inequality~\eqref{Azuma} gives
\[ \Pr(|\tilde{X} - \E(\tilde{X})| \geq n/2^{6 \log^a n}) \leq 2\exp\left\{ -\frac{n^2}{2 n^{3/2} 2^{12 \log^a n}}\right\}. \]
Since there is almost no difference between the original process and the modified process, we obtain the following bound
on the probability that the number $X$ of nodes with their interactions counts within $\delta \log^{1-a} n$
from $2(t'-t)/n$ is less than $n - 3n/2^{6\log^{a} n}$.
Recall that both $\E(X)$ and $\E(\tilde{X})$ are at least $n - n/2^{6\log^{a} n}$.
\begin{eqnarray}
\Pr( X < n - 3n/2^{6\log^{a} n})
   & \le & \Pr( |X - \E(X)| \ge 2n/2^{6\log^{a} n}) \nonumber \\
   & \le & \Pr( |\tilde{X} - \E(\tilde{X})| + |\tilde{X} - X| + |\E(\tilde{X}) - \E({X})| \ge 2n/2^{6\log^{a} n})
         \nonumber \\
   & \le & \Pr( |\tilde{X} - \E(\tilde{X})| \ge n/2^{6\log^{a} n}
               \; \mbox{or} \; X \neq \tilde{X}) \nonumber \\
   & \le & \Pr( |\tilde{X} - \E(\tilde{X})| \ge n/2^{6\log^{a} n}) + \Pr(\overline{\cE}) \nonumber \\
   & \le & \exp\{ - \Theta(\sqrt[4]{n})\}.  \label{lpj72la}
\end{eqnarray}

To conclude the proof of the claim, we note that provided that the local times of the nodes remain consistent
(a high-probability event; see Lemma~\ref{lem_concentration_simple}),
the $\epochstep$ counter of each node in $W$ with the number of interaction within $\delta \log^{1-a} n$
from $2(t'-t)/n$ is within $c\log^a n + \delta \log^{1-a} n \le (c/2)  \log^{1-a} n$ from
$2(t'-t)/n$.
For the last inequality and for~\eqref{hjw7aq} we take $\d$ satisfying
$5C^{1/2} \le \d \le c/3$, which is possible for sufficiently large $C$ and $c = C^{3/4}$.
\end{proofof}

%

The proof of Lemma~\ref{Lem:phaseInv} uses the analysis of one cancellation stage
(Lemmas\ref{cancellations} for a phase $p(j,i) < p_c$ and
Lemma~\ref{cancellations_small} for the phase $p_c$) and the analysis of
the subsequent doubling stage
(Lemma~\ref{Lem:phaseduplications}).


\begin{sloppy}

\begin{lemma}
\label{cancellations}
Consider epoch $j$ and phase $i$ such that $p(j,i) < p_c$
and assume that
the condition $\EpochInv(j)$ holds at a (global) time step $t$ and
the condition $\PhaseInvA(j,i)$ holds at the step $t_i$ (defined in Lemma~\ref{Lem:phaseInv}).
%
That is, in particular, at step $t_i$
the set $W$ of nodes which are normal and in the beginning part,
that is, within the first $c\log^{1-a} n$ steps,
of the cancellation stage
of phase~$i$ in epoch~$j$ has size
at least $n(1 - (i+1)/2^{2\log^{a}n})$.
Then \whp at step $t'_i = t_i + (C/4 - c/2)n\log^{1-a} n$ the following conditions hold.
\begin{enumerate}
\item\label{ghdop1}
The set $W'$ of
normal nodes in the end part (within the last $c\log^{1-a} n$ steps)
of this cancellation stage has size
at least $n(1- (i+1 +o(1))/2^{2\log^a n})$.
\item\label{ghdop2}
The total value, \wrt\ the end of epoch $j$, of the tokens in $U' = V \setminus W'$ is at most
\mbox{$n (i+1+o(1))/2^{2\log^{a}n}$}.
\item\label{ghdop3}
There are at least $(6/10)n$ empty nodes.
\end{enumerate}
\end{lemma}

\end{sloppy}

\begin{proof}
For the phase  $p(j,i) < p_c$,
the difference between the total value of the $A$ tokens and $B$ tokens
\wrt (the beginning of) this phase is $2^{p(j,i)} |a_0 - b_0| \le n/3$.

We consider the set $W$ of the
normal nodes which are at step $t_i$
in the beginning of the cancellation stage of phase $i$ (in epoch $j$)
and the set $\cT$
of tokens in these nodes.
All tokens in $\cT$ have value $1/2^{p(j,i)}$ and
we analyze how they cancel out each other in steps $t_i, t_i+1, \ldots, t'_{i}$.

We assume
first that tokens from $\cT$ cancel out (and are removed from $\cT$) when, and only when,
two opposite-type tokens from $\cT$ interact.
We note that in the actual process, these two tokens might not cancel out, if one of them is already
outside this cancellation stage, or one of them has canceled out earlier with a token not in $\cT$.
We view the sequence of steps $t_i, t_i+1, \ldots, t'_i$ as a sequence of
$(C/4 - c/2)\log^{1-a} n$ periods, with each period consisting of
$n$ interactions.
We consider one of these periods, denote by $z$
the smaller of the count of $B$ tokens and the count of  $A$ tokens in $\cT$
at the beginning of this period,
and show that if $z \ge n/40$, then
\whp
at the end of the period
the smaller of the count of $B$ tokens and the count of $A$ tokens in $\cT$ is at most
$(41/42)z$.

We call an interaction a success, if two tokens from $\cT$ cancel out or
the number of $B$ tokens in $\cT$ or the number of $A$ tokens in $\cT$
is already less than $(41/42) z$.
Thus the probability that a given interaction in this period
is a success is at least $2\cdot ((41/42)z/n)^2$, so
the number of successes is at least $2\cdot ((41/42)^2 (z/n) z \ge (2/42)z$ in expectation
and at least $z/42$ \whp (from the Chernoff bound~\eqref{ChernoffC}).
The event that there are
at least $z/42$ successes
implies that at least $z/42$ pairs of tokens in $\cT$ have canceled out, so
the smaller of the count of $B$ tokens and the count of $A$ tokens remaining in $\cT$
at the end of the period is at most  $(41/42)z$.
Therefore \whp after a sufficiently large constant number of periods, the
smaller of the count of $B$ tokens and the count of $A$ tokens remaining in $\cT$
is at most $n/40$.

An interaction of two opposite-type tokens from $\cT$
in one of the steps $t_i, t_i+1, \ldots, t'_{i}$ is not a cancellation,
if one of the two tokens is already beyond this cancellation stage
or it has canceled out earlier with a token of value $1/2^{p(j,i)}$ not in $\cT$.
Claim~\ref{claim_close} implies that \whp the size of the set $W_1 \subseteq W$
of nodes in $W$ which move beyond
this cancellation stage by step $t'_i$ is $O(n/2^{6 \log^a n})$, since
the $\epochstep$ counters of these nodes at step $t'_i$ are greater than
$(iC + C/2)\log^{1-a} n  = 2(t'_i - t)/n + (c/2)\log^{1-a} n$.
From the assumptions of the lemma
(from the condition $\PhaseInvA(j,i)$  at step $t_i$),
the number of tokens of value $1/2^{p(j,i)}$ which appear in steps
$t_i, t_i+1, \ldots, t'_{i}$ but are not in $\cT$ is at most $n (i+1)/2^{2\log^{a}n}$,
since these tokens originate from tokens which were in $U = V\setminus W$
at step $t'_i$.
Thus \whp the number $X_B$ of $B$-tokens or the number $X_A$ of $A$-tokens remaining in $\cT$
at step $t'_i$
is at most $(1/40 + o(1))n$.

We bound now the total number of tokens at step $t'_i$.
Let $\cF$ denote the set of tokens at step $t'_i$ other than the tokens remaining in $\cT$,
and
let $\cF(q)$ and $\cF_A(q)$ (resp.\ $\cF_B(q)$)
denote the total value of all tokens in $\cF$ and the total value of all $A$ tokens (resp.\ all $B$ tokens)
in $\cF$ w.r.t.\ phase $q$ (of epoch $j$).
All tokens in $\cF$ originate from tokens which were at step $t_i$ in
$U = V\setminus W$ or in $W_1$, so their total value $\cF(\log^a n)$
w.r.t.\ the end of epoch $j$ is at most
$n (i+1)/2^{2\log^{a}n}$
(from tokens in $U$ at step $t_i$, since $\PhaseInvA(j,i)$ holds at step $t'_i$)
plus  $2^{\log^a n} \cdot O(n/2^{6 \log^a n})$ (from tokens in $W_1$ at step $t_i$;
see Claim~\ref{claim_close}).
Thus
\begin{equation}\label{jkcv8a2s}
  \cF(\log^a n) \le  n (i+1+o(1))/2^{2\log^{a}n} = o(n).
\end{equation}

For the number $X_B$ of $B$-tokens in $\cT$ at step $t'_i$, we have
\[ X_B \le X_B + \cF_B(i) < X_A + \cF_A(i) \le X_A + \cF_A(\log^{a} n),
\]
where  $X_B + \cF_B(i)$ (resp.\ $X_A + \cF_A(i)$)
is the total value of $B$ tokens (resp.\ $A$ tokens) at step $t'_i$ and
the inequality $X_B + \cF_B(i) < X_A + \cF_A(i)$ follows from $b_0 < a_0$.
Thus at step $t'_i$, $X_B \le X_A + o(n)$ and
we showed earlier that
$X_B \le (1/40 + o(1))n$ or
$X_A \le (1/40 + o(1))n$, so
in either case
\begin{equation}\label{kld90a}
  X_B \le (1/40 + o(1))n.
\end{equation}

At step $t'_i$, the difference between the total value of $A$ tokens and $B$ tokens w.r.t.\ phase $i$
is equal to
\begin{equation}\label{jk38as}
 (X_A + \cF_A(i)) - (X_B + \cF_B(i)) \; = \; 2^{p(j,i)} |a_0 - b_0|  \; \le \; \frac{n}{3}.
\end{equation}
The total values on the left-hand side above are calculated for step $t'_i$, but the difference
is invariant throughout the whole protocol. The inequality holds because we
consider phase $p(j,i) < p_c$.
Summarizing, the number of tokens at step $t'_i$ is \whp at most
$X_A + X_B + \cF(\log^a n)$ and
from~\eqref{jkcv8a2s}--\eqref{jk38as},
\[
X_A + X_B + \cF(\log^a n)
 \; \le \; \frac{n}{3} + 2X_B + \cF_B(i) - \cF_A(i) + \cF(\log^a n)
 \; \le \; \frac{n}{3} + 2X_B + 2\cF(\log^a n)
 \; \le \; \frac{4}{10}n.
\]
Therefore there are at least $(6/10)n$ empty nodes at step $t'_i$; the claim~\ref{ghdop3}
of the lemma.

Let $W_2\subseteq W$ be the set of nodes in $W$ which at step $t'_i$ have not reached
yet the end part of the cancellation stage.
The tokens at step $t'_i$ which are not normal (that is, are out-of-sync)
or not in the end part of the cancellation stage
of phase $i$
originate from tokens which were in $U \cup W_1 \cup W_2$ at step $t_i$.
\Wrt the end of the epoch, the total value
of the tokens originating from $U$  is at most $n (i+1)/2^{2\log^{a}n}$
(condition $\PhaseInvA(j,i)$ at step $t_i$)
and the total value of the tokens in $W_1\cup W_2$
is at most
$2^{\log^a n} \cdot O(n/2^{6 \log^a n})$ (from Claim~\ref{claim_close}).
Thus the total value, \wrt the end of the epoch,
of the tokens at step $t'_i$ which are not normal
or not in the end part of the cancellation stage
of phase $i$ is at most $n (i+1+o(1))/2^{2\log^{a}n}$; the claim~\ref{ghdop2} of the lemma.

The nodes which at step $t'_i$ are normal and in the end part of the cancellation
stage of phase~$i$ are all nodes other than
({\it i\/}) the nodes with tokens which are not normal
or not in the end part of this cancellation stage, and
({\it ii\/})
the nodes without token and not in the end part of this cancellation
stage.
There are at most $n (i+1+o(1))/2^{2\log^{a}n}$ nodes of type ({\it i\/})
(from the conclusion~\ref{ghdop2} of the lemma) and
at most $O(n/2^{6 \log^a n})$ nodes  of type ({\it ii\/}) (from Claim~\ref{claim_close}),
implying the claim~\ref{ghdop1} of the lemma.
\end{proof}


\begin{lemma}
\label{cancellations_small}
Consider epoch $j$ and phase $i$ such that $p(j,i) = p_c$
and assume that
the condition $\EpochInv(j)$ holds at a (global) time step $t$ and
the condition $\PhaseInvA(j,i)$ holds at the step $t_i$ (defined in Lemma~\ref{Lem:phaseInv}).
Then \whp at step $t'_i = t_i + (C/4 - c/2)n\log^{1-a} n$
the total value of the minority-opinion tokens
\wrt the end of epoch $j$ is $O(n\log n/2^{\log^{a}n})$.
\end{lemma}
\begin{proof}
The proof is similar to the proof of Lemma~\ref{cancellations} and we use the same terminology as in that proof.
In particular, we divide the time interval $[t_i, t'_i]$ into
periods of $n$ interactions and
we refer to the terms $\cT$, $z$, $\cF$ and $\cF(q)$ defined in the proof of Lemma~\ref{cancellations}.
Now, however, since
the difference between the total value of the majority-opinion $A$ tokens and the minority-opinion $B$ tokens w.r.t.\ phase $i$
is at least $n/3$ (by the definition of phase $p_c$),
the number of the $A$ tokens in $\cT$ remains greater than
the number of the $B$ tokens in $\cT$
by at least $n/3 - o(n) \ge n/4$.
Thus, considering one period of $n$ interactions,
$z$ is now the number of $B$ tokens in $\cT$ at the beginning of this period
(since now guaranteed to be smaller than the number of the $A$ tokens in $\cT$),
and we show that if $z \ge n/2^{\log^a n}$, then \whp
the number of $B$ tokens in $\cT$ is at most $(4/5)z$ at the end of the period.

The probability that a given interaction in the current period
is a success (that is, a meeting of two opposite-type tokens from $\cT$ or the number of $B$ tokens in $\cT$ already smaller then
$(4/5)z$)
is at least $2\cdot (4/5)z/n \cdot 1/4 = (2/5)z/n$.
This implies that
the number of successes is at least $(2/5)z$ in expectation
and at least $z/5$ \whp  (from the Chernoff bound~\eqref{ChernoffC}).
The event that there are
at least $z/5$ successes in this period
implies that at least $z/5$ $B$-tokens in $\cT$ have canceled out during this period, so
at most $(4/5)z$-$B$ tokens remain in $\cT$
at the end of the period.
Thus \whp after $\Theta(\log^{a} n)$ periods,
the number of $B$-tokens remaining in $\cT$ is at most $n/2^{\Theta(\log^{a} n)}$.

As in the proof of Lemma~\ref{cancellations}, we might have overcounted the number of cancellations of $B$
tokens in $\cT$, but not more than by $O(n/2^{6 \log^a n}) + n (i+1)/2^{2\log^{a}n} = n (i+1+o(1))/2^{2\log^{a}n}$.
Therefore the total value of the $B$ tokens at step $t'_i$ \wrt the end of the epoch is \whp at most
\[  n (i+1+o(1))/2^{2\log^{a}n} \cdot 2^{\log^{a} n} + \cF(\log^a n) \; = \;
        O(n \log n /2^{\log^{a}n}).
\]
\end{proof}


\begin{sloppy}

\begin{lemma}\label{Lem:phaseduplications}
Consider epoch $j$ and phase $i$ such that $p(j,i) \le p_c$
and assume that at the (global) time step~$t$ the condition $\EpochInv(j)$ holds
and at step $t'_i = t_i + (C/4 - c/2)n\log^{1-a} n$ (where $t_i$ defined in Lemma~\ref{Lem:phaseInv})
the conditions~\ref{ghdop1}--\ref{ghdop3} of Lemma~\ref{cancellations} hold.
Then \whp at step $t''_i =  t_i + (C/2 - c/2)n\log^{1-a} n$
the following conditions hold.
\begin{enumerate}
\item\label{kl58xna2a}
Let $W''$ be the set of nodes $v$ such that
$v$ is normal and in the end part (within the last $c\log^{1-a} n$ steps)
of phase $i$ in epoch $j$, and if $v$ contains a token, then
its value is $1/2^{p(j,i) + 1}$ (as expected at the end of this phase).
The size of $W''$ is at least $n(1- (i+1 +o(1))/2^{2\log^a n})$.
\item\label{kl52a}
The total value, w.r.t.\ the end of epoch $j$, of the tokens in $U'' = V \setminus W''$ is at most
\mbox{$n (i+1+o(1))/2^{2\log^{a}n}$}.
\end{enumerate}
\end{lemma}

\end{sloppy}

\begin{proof}
For the step $t'_i$,
let $W'$ and $U'$ be the sets of nodes defined in
the conditions~\ref{ghdop1}--\ref{ghdop2} of Lemma~\ref{cancellations}.
In particular, $W'$ is the set of nodes which, step $t'_i$, are normal and
in the end part of the cancellation stage of phase $i$ (in epoch $j$).
Let $\cT$ denote the set of tokens in $W'$ and $\cF$ the set of tokens in $U'$.
There are at most $(4/10)n$ tokens in $\cT$ (from the condition~\ref{ghdop3}
of Lemma~\ref{cancellations}) and all of them have value $1/2^{p(j,i)}$.
At step $t''_i$ most of the nodes should be in the end part of phase $i$ (from Claim~\ref{claim_close}).
We will show that by this step $t''_i$ \whp all but $O(n/2^{6\log^a n})$ tokens in $\cT$ split to
half tokens of value $1/2^{p(j,i) + 1}$.
This will imply that at step $t''_i$ \whp the total value (\wrt to the end of the epoch)
of the tokens which are not normal
or have value different than $1/2^{p(j,i) + 1}$ or are in nodes not in the end part of phase $i$
(that is, the the total value of the tokens in $U''$)
is at most
$n (i+1+o(1))/2^{2\log^{a}n}$ (the contribution from the tokens originating from tokens in $\cF$)
plus $2^{\log^a n} \cdot O(n/2^{6\log^a n})$
(the contribution from the tokens in  $\cT$ which have not split)
plus $2^{\log^a n} \cdot O(n/2^{6\log^a n})$
(the contribution from the nodes in $W'$ which at step $t''_i$
are not in the end part of phase $i$; using Claim~\ref{claim_close}),
so at most $n (i+1+o(1))/2^{2\log^{a}n}$; the claim~\ref{kl52a} of the lemma.
The claim~\ref{kl58xna2a} of the lemma follows by observing
that each node $v$ from $W'$ belongs to $W''$ unless $v$ is a node in $U''$ with a token
(the claim~\ref{kl52a} of the lemma implies that
\whp there are at most $n (i+1+o(1))/2^{2\log^{a}n}$ such nodes)
or $v$ is not in the end part of the phase $i$ at step $t''_i$
(Claim~\ref{claim_close} implies that there are $O(n/2^{6\log^a n})$ such nodes).
It remains to show that  by step $t''_i$ \whp all but $O(n/2^{6\log^a n})$ tokens in $\cT$ split.

The assumptions of the lemma, Claim~\ref{claim_close} and Lemma~\ref{lem_concentration_simple}
imply that \whp\ in each of the steps $t'_i, t'_i+1, \ldots, t''_i$
there are at most $(4/5 +o(1))n$ tokens:
$2x + o(n)$ tokens obtained from doubling, and possible subsequent further splitting, of
$x$ tokens from $\cT$;
at most $(4/10)n - x$ tokens remaining in $\cT$, for some $x \le (4/10)n$;
and at most  $n (i+1+o(1))/2^{2\log^{a}n}$ tokens originated from $\cF$.
Thus \whp\ there are at least $(1/5 - o(1))n$ empty nodes in each of these steps.
Let
$t'''_i = t_i + (C/4 + c/2)n\log^{1-a} n$.
At this step $t'''_i$ most of the nodes should be in the beginning part of the doubling stage of phase $i$.
We consider interactions in steps from $t'''_i$ to $t''_i$, when most of the nodes
are in the doubling stage of phase $i$.

We assume first that a token in $\cT$ splits whenever it interacts with an empty node
(the original token is then removed from $\cT$).
In the actual process, the splitting would not happen, if one or both of the nodes are
outside this doubling stage.
We view the sequence of steps $t'''_i, t'''_i+1, \ldots, t''_i$ as a sequence of
$(C/4 - c)\log^{1-a} n$ periods, with each period consisting of
$n$ interactions.
For each of these periods, if there are $z \ge n/2^{\log^{1-a}n}$ tokens in $\cT$
at the beginning of the period, then
\whp at the end of the period at most $(3/4)z$ tokens remains in $\cT$.
Indeed, the probability that a given interaction in this period
matches one remaining token in $\cT$ with an empty node
is at least $2\cdot((3/4)z/n)\cdot(1/5 - o(1))$,
unless the number of tokens remaining in $\cT$ is already less than $(3/4)z$.
Thus the expected number of successful interactions in this period
(interactions when a token in $\cT$ splits or the number of tokens in $\cT$ is less than $(3/4)z$)
is at least $(3/10 - o(1))z$, so the actual number of successful interactions is
\whp at least $z/4$ (from Chernoff bound~\eqref{ChernoffC}).
This implies that the size of $\cT$ reduces \whp to less than $(3/4)z$ in this period.
Therefore after the $(C/4 - c)\log^{1-a} n$ periods (that is, at step $t''_i$)
\whp the number of tokens remaining in $\cT$ is less than $n/2^{\log^{1-a} n}$,
for sufficiently large $C$.

Some of the interactions counted above as splitting tokens in $\cT$ did not do so,
if one of the nodes was not in the current doubling stage.
Since \whp there were at most $O(n/2^{6\log^a n})$ nodes which were not in this doubling stage
throughout the steps $t'''_i, t''_i+1, \ldots, t''_i$ (from Claim~\ref{claim_close}),
the number of tokens remaining in $\cT$ at step $t''_i$
is at most $n/2^{\log^{1-a} n} + O(n/2^{6\log^a n}) = O(n/2^{6\log^a n})$.
\end{proof}


\begin{proofof}{Lemma~\ref{lem_concentration_simple}}
For the number $Z_v$ of interactions of node $v$, it can be shown using
the Chernoff bound~\eqref{ChernoffA}, that
\begin{eqnarray}
\Pr(|Z_v - 2t/n| \ge c\log n)
 &\leq& \exp\{- (\sqrt{C}/16) \cdot \log n\}.\label{klwe5a}
\end{eqnarray}
The probability on the left-hand side  above is maximized for $t = 2Cn\log n$.
For this value of $t$, we get the right-hand side of~\eqref{klwe5a} by applying~\eqref{ChernoffA} with
$\ep = c/(4C)$.
Using Inequality~\eqref{klwe5a} and the union bound over all nodes, we conclude that
for a sufficiently large $C$,
$\mbox{\whp}(C)$, for each node $v$, $|Z_v - 2t/n| \le c\log n$.
\end{proofof}

Lemmas \ref{lem_broadcast_simple} and~\ref{lem_broadcast} describe the performance of the broadcast process in
the population-protocol model.
At the beginning of the broadcast process (step $0$),
at least one node
has a message.
In each step two nodes chosen uniformly at random interact
and whenever an informed node $u$ (i.e., a node which already has the message) interacts with
an uniformed node $v$, then
$v$ gets the message.
An uninformed node may become informed during its current interaction, even if the other
node is also uninformed.
The informed nodes do not lose the message.
In our \FastExactMajorityA protocol, we use the broadcast protocol to move the nodes to the next epoch.
The broadcast is initiated when
the first node which reaches the beginning of the next epoch $j+1$ initiates the broadcast of the message that
each node has to move to epoch $j+1$.
The nodes move to epoch $j+1$ when they receive the message, but may also progress to that epoch
when they reach by themselves the end of epoch $j$.

\begin{proofof}{Lemma~\ref{lem_broadcast}}
From step $t_1$, consider the sequence of periods, with each period consisting of
$n$ consecutive interactions (steps).
First the number of informed nodes geometrically increases from period to period until it reaches $n/2$.
Then the number of uninformed nodes geometrically decreases until it drops below $n/2^{6\log^b n}$.
More specifically, \whp the following two statements hold.
\begin{enumerate}
\item
In each period which starts with fewer than $n/2$ informed nodes,
the number of informed nodes
increases at least by factor $3/2$ or to $n/2$.
\item
In each period which starts with the number of informed nodes at least $n/2$ but less than
$n-n/2^{6\log^b n}$,
the number of uninformed nodes decreases at least by factor $2/3$ or to
$n/2^{6\log^b n}$.
\end{enumerate}

In order to show the first statement,
consider period $\t$ such that
the number $I(\t)$ of informed nodes at the beginning of this period is less than $n/2$.
We say that a given interaction in this period is a success,
if a new node becomes informed or the number of informed nodes is already at least $n/2$.
Let $X$ denote the number of successes.
If $X \ge I(\t)/2$, then the number of informed nodes
increases in this period at least to $(3/2)I(\t)$ or $n/2$.
For a given interaction, if the number of informed nodes is already at least $n/2$, then
the probability of success is $1$.
If the number of informed nodes is still less than $n/2$,
then the probability of success is at least $I(\t)/n$.
Indeed,
if the first node selected for this interaction is uninformed,
then
success comes when the second selected node is an informed node, so
with probability at least $I(\t)/n$.
If the first  selected node is an informed node,
then success comes when the second selected node is uninformed, so
with probability at least $1/2 > I(\t)/n$.
Thus in all cases the probability of success in a given interaction is at least $I(\t)/n$,
so $\E(X) \ge I(\t)$ and the Chernoff bound~\eqref{ChernoffA} gives
\[\Pr(X \leq I(\t)/2) \leq \exp\left\{ -\left(\frac{1}{2}\right)^2 \frac{1}{4}\,{I(t)} \right\} = 1/n^{\omega(1)} . \]

To show the second statement, consider a period $\t$ such that at the beginning of this period
the number of informed nodes is at least $n/2$ but less than
$n-n/2^{6\log^b n}$. Equivalently,
the number $U(\t)$ of uninformed nodes is at most $n/2$
but still greater than $n/2^{6\log^b n}$.
An interaction is a success, if a new node is informed or the number of uninformed nodes
is at most $\max\{U(\t)/2, n/2^{6\log^b n}\}$.
If the number $X$ of successes is at least $U(\t)/3$, then
the number of uninformed nodes decreases in this period at least to $(2/3)U(\t)$ or
$n/2^{6\log^b n}$.
Similarly to above, the probability of success in a given interaction is at least $U(\t)/(2n)$,
so $\E(X) \ge U(\t)/2$ and the Chernoff bound~\eqref{ChernoffA} gives
\[\Pr(X \leq U(\t)/3) \leq \exp\left\{ -\left(\frac{1}{6}\right)^2 \frac{1}{4} \, \frac{U(\t)}{2} \right\} = 1/n^{\omega(1)} . \]

The two statements imply Condition 1 of the lemma: with probability at least $1 - n^{-\omega(1)}$,
all but at most $2n/2^{6\log^b n}$ nodes become informed
within $c_1 \log^b n$ periods from step $t_1$,
that is, in steps $\{ t_1+1, t_1+2, \dots , t_2 \}$, for $c_1 \ge 24$.
($n/2^{6\log^b n}$ nodes were informed before the step $t_1$ and at most $n/2^{6\log^b n}$
nodes will be informed after the step $t_2$.)
Condition 2 follows immediately from Lemmas~\ref{lem_concentration_simple} and~\ref{lem_broadcast_simple}.

In order to show Condition 3, we consider any sequence of $c_1 n \log^b n$ consecutive interactions,
and show that \whp $n -O(n/2^{\log^b n})$ nodes interact
at least $c_1\log^b n$ times but not more than $3c_1\log^b n$ times.
Denoting $\mu = 2c_1\log^b n$ the expected number of interactions per node
and using the Chernoff bound~\eqref{ChernoffA},
we get the following bound on the probability that the number $Z_u$ of interactions of a node $u$ deviates from $\mu$ by more
than $c_1\log^b n$.
\[ \Pr( |Z_u - \mu| \ge c_1 \log^b n) \leq \exp\left\{ -\frac{c_1 \log^b n}{8}\right\}. \]
Thus for sufficiently large $c_1$,
the expected number of nodes which interact at least $c_1 \log^b n$ and at most $3c_1 \log^b n$ times
is at least $n(1-1/2^{6\log^b n})$.
By applying the same techniques as in the proof of Claim \ref{claim_close}
(that is, by constructing an appropriate martingale and deriving
a bound analogous to~\eqref{lpj72la}),
we obtain an upper bound of $n^{-\omega(1)}$ on the probability that
the number of nodes which interact between $c_1 \log^b n$ and $3c_1 \log^b n$ times
is less than  $n - 3n/2^{6\log^b n}$.
\end{proofof}

\subsection{Pseudocodes and further details for Section \ref{Sec-FastOptStates}
-- protocol \FastExactMajorityB}

During a clock-clock interaction (Algorithm~\ref{FastB-clock-clock}),
 the nodes first check the consistency of their $\epochstep$ counters
and switch to $\fail$ states, if the difference between the counters modulo $2C\log n$ is greater than
$(1/4)C\log n$.
If the counters are consistent, then they both are incremented by $1$.
If one of the counters is in the final quarter
of the range while the other in the first quarter, then the former is reset to $0$.
This is the mechanism of pulling up to the next epoch the clock nodes lagging behind.

During the interaction between a worker node $v$ and a clock node $u$ (Algorithm~\ref{FastB-worker-clock}),
the worker node first uses the time reading from the clock node to decide whether to progress its computation to the
next stage, phase or epoch, and then executes
the procedure $\nextstep\_\mbox{\it normal}(v)$ or $\nextstep\_\mbox{\it outofsync}(v)$, depending on the type of
the worker node. These two $\nextstep$ procedures are as defined in \FastExactMajorityA.
The clock node $u$ does not change its state during this interaction.

\begin{algorithm}[t]
\lIf{$\neg \consistent(v.\Time, u.\Time)$}{
    $v.\fail, u.\fail \Let \true$;
}
\uElse{
                \lFor{each $x \in \{ u,v\}$}{
                     $x.\epochstep \Let x.\epochstep + 1$;
                  }
               let $v.\epochstep \le u.\epochstep$; \tcp{the other case is analogous}
              \uIf{$((v.\epochstep \le (1/4)C\log n) \wedge (u.\epochstep \ge (7/4)C\log n))$}{
              \tcp{assuming $u$ in the final part of an epoch and $v$ in the next epoch}
             $u.\epochstep \Let 0$;
             }
}
 \caption{\FastExactMajorityB protocol -- a clock-clock interaction\label{FastB-clock-clock}}
\end{algorithm}

\begin{algorithm}[t]
let $v$ be a worker node and $u$ a clock node;\;
\tcp{set the default ``no progress'' for the $\nextstep(.)$ procedures}
``$v.\phasestep = 0$'' $\Let \false$;\;
``$v.\epochstep = 0$'' $\Let \false$;\;
\;
\uIf{$v.\normal$}{
    \uIf{$v.\epochpart = 0$}{
      \uIf{$(v.\ageinepoch < u.(\epochpart,\phase)) \wedge (\mbox{$u$ is not in the end of epoch})$}{
          \tcp{$v$ moves to the next phase}
          $(v.\epochpart,v.\phase) \Let (v.\epochpart,v.\ageinepoch) +1$;
          \tcp{$v.\phase$ only for $\nextstep(.)$, not for storing}
          $v.\stage \Let \beginningphase$;\;
          ``$v.\phasestep = 0$'' $\Let \true$;
       }
       \uElseIf{$(v.\ageinepoch=u.\phase)$}{
          \lIf{$u.\stage$ is later than $v.\stage$}{advance $v.\stage$ to the next stage;}
       }
    }

    \uElseIf{$(v.\epochpart = 1) \wedge (\mbox{$u$ is in the beginning of epoch})$}{
        \tcp{$v$ moves to the start of the next epoch}
        $v.(\epoch,\epochpart) \Let v.(\epoch,\epochpart) + 1$;\;

    }
    $\nextstep\_\mbox{\it normal}(v)$;
    }
\uElse{
    \tcp{$v$ has an out-of-sync token}
    \uIf{$(v.\epochpart = 0) \wedge (u.\epochpart = 1) \wedge
           (\mbox{$u$ is not in the end of epoch})$}{
        \tcp{$v$ moves to the second part of epoch}
        $v.\epochpart \Let 1$;\;
        ``$v.\phasestep = 0$'' $\Let \true$;\;
    }
    \uElseIf{$(v.\epochpart = 1) \wedge (\mbox{$u$ is in the beginning of epoch})$}{
        \tcp{$v$ moves to the start of the next epoch}
        $v.(\epoch,\epochpart) \Let v.(\epoch,\epochpart) + 1$;\;
        ``$v.\epochstep = 0$'' $\Let \true$;\;
    }
    $\nextstep\_\mbox{\it outofsync}(v)$;
}
 \caption{\FastExactMajorityB protocol -- a worker-clock interaction\label{FastB-worker-clock}}
\end{algorithm}

\begin{sloppy}

The worker-worker interactions (Algorithm~\ref {FastB-worker-worker}) are essentially the same
as in the \FastExactMajorityA protocol (Algorithms~\ref{AlgoNormalProgress} and~\ref{AlgoOutofsync})
but now without the updates involving step counters.

\end{sloppy}

The exact-majority protocol in~\cite{DBLP:conf/soda/AlistarhAG18}
relies on the leaderless phase clock which \whp\ keeps the step counters of {\em all\/} clock nodes synchronized within
an interval of length $c\log n$, where $c$ is constant considerably smaller than the constant $C$.
In our protocol, we need the clock nodes to stay synchronized within an interval of length $\Theta(\log^{1-a}n)$.
This cannot be achieved ``\whp'', but we can show that only $n/2^{\Theta(\log^a n})$ clocks fall outside of such tight
synchronization.
The synchronization of the clock nodes is described
in the following lemma, which can be proven using Claim~\ref{claim_close} and Lemma~\ref{lem_broadcast}.
We denote by $n_c$ and $n_w$ the number of clock nodes and the number of worker nodes, respectively,
and assume that both are $\Theta(n)$.

\begin{lemma}
Assume that the counter of each clock node is at most $c\log n$ or
greater than $(7/4)C\log n$, and at least $n_c(1-1/2^{6\log^a n})$ clock nodes have
their counter at most $C\log^{a} n$.
Call these conditions the $\EpochInv\_\mbox{\it Clocks}$.
Then \whp\
\begin{enumerate}
\item
at the beginning of each of the subsequent $(7/8)C\log n$ periods of $n$ interactions each,
$n_c(1-1/2^{5\log^a n})$ clock nodes have counters within $c\log^{1-a} n$ of each other, and
\item
within additional $(1/4)C\log n$ periods
the $\EpochInv\_{\mbox Clocks}$ condition holds again.
\end{enumerate}
\end{lemma}

This lemma implies that
during each period of $n$ interactions, \whp\ only
$n/2^{\delta_1\log^a n}$ worker nodes interact with desynchronized
clock nodes, for some constant $\delta_1 > 1$.
This in turn implies that for some constant $1 < \delta_2 < \delta_1$,
$n_w - n/2^{\delta_2\log^a n}$ worker nodes,
progress orderly through all canceling/doubling phases of an epoch
as in the \FastExactMajorityA protocol.

\begin{sloppy}

The invariant at the beginning of an epoch $j$ is the $\EpochInv\_\mbox{\it Clocks}$ condition together
with the following
$\EpochInv\_\mbox{\it Workers}(j)$.
\begin{enumerate}
\item
At least $n_w(1 - 1/2^{3\log^{a}n})$ worker nodes are in normal states, in epoch $j$, phase $0$
and with $\stage = \beginningphase$.
\item
For each remaining worker node $u$,
\begin{enumerate}
\item
$u$ is in a normal state in the second part of epoch $j-1$, or
\item
$u$ is in a normal or out-of-sync state in epoch $j$ and $u.\ageinepoch \le c\log^{1-a} n$.
\end{enumerate}
\end{enumerate}
Lemma~\ref{Lem:epochinv} holds for the \FastExactMajorityB protocol with the same wording,
but with the $\EpochInv(j)$ condition
modified as above.
This lemma can be proven for the \FastExactMajorityB protocol by closely following the proofs of
Lemmas~\ref{Lem:phaseInv}-\ref{Lem:phaseduplications}. The underlying premise is the same:
\whp\ $n_w(1 - 1/2^{\Theta(\log^{a}n)})$ worker nodes move through the phases of an epoch in
synchronized manner, having the number of interactions close to the expectation.
The remaining $O(n_w/2^{\Theta(\log^{a}n)})$ out-of-sync workers have enough attempt at splitting
to ensure that \whp\ all tokens split to the values required at the end of the epoch.

\end{sloppy}

%
%
%
%

It remains to show that the \FastExactMajorityB protocol can be initialize,
so that we have linear-size sets of worker and clock
nodes and their states satisfy the $\EpochInv(0)$ condition.
Initially, each node has a token, either $A$ or $B$, of the initial value $1$, and all nodes are declared as workers.
We make the first phase special by allowing it to run for $C\log n$ steps.
The nodes count the steps of this phase by themselves, but the required $\Theta(\log n)$ states for this counting
will be reused in the subsequent computation, so we total state count stays within $\Theta(\log n)$.
We allow in this phase only the following operations on tokens.
If two value-$1$ tokens of opposite type interact, then they cancel out, and if this is the first
interaction for each of the two nodes, then one node, say the node which has had token $B$,
becomes a clock node
while the other node becomes permanently fixed as a worker node.
If two value-$1$ tokens of the same type interact and their step counters have different parity,
then the tokens are combined into one token of value $2$.
The combined token is taken by one node, say by the node with even step counter,
which becomes permanently fixed as a worker node, while the other node becomes a clock node.

It can be shown that if the initial size of the minority opinion is at least $n/4$, then \whp
$\Theta(n)$ clock nodes are created and $\Theta(n)$ nodes are fixed as workers, since each of the first
$n/8$ interactions has a positive constant probability of being a cancellation,
creating one clock and fixing one node as a permanent worker.
If the size of the minority opinion is less than $n/4$, then \whp\ we get
$\Theta(n)$ clock nodes and $\Theta(n)$ permanent workers by combining majority tokens.

The first node which reaches the end of the initial special phase initiates a broadcast
to move the system \whp\ into a global configuration which satisfies conditions analogous to $\EpochInv(0)$, but with the
{\em base value} of tokens changed from $1$ to $2$.
We will have in the beginning of phase $0$ a mix of tokens of values $2$ and $1$. To deal with this
we set the flag $\doubled$ for the tokens of value $1$.

Lemma~\ref{Lem:epochinv} adapted to the \FastExactMajorityB protocol and the above discussion
leads to our final result stated in Theorem~\ref{thm-xxxx},
 where the ``in expectation'' part can be argued as in~\cite{DBLP:conf/soda/AlistarhAG18}.

\

\begin{algorithm}
\uIf{$v.\normal \wedge u.\normal$}{
\uIf{$((v.\epoch = u.\epoch) \wedge (v.\epochpart = u.\epochpart = 0) \wedge (v.\ageinepoch= u.\ageinepoch))$}{
\uIf{$v$ and $u$ have opposite tokens and are in the canceling stage}{
            $u.\token \Let \notoken$;~ $v.\token \Let \notoken$\;
}
\uElseIf{exactly one of $v$ and $u$ has a token and both nodes in the doubling stage}{
        say $v$ is the node with a token;\;
        \lIf{$\neg v.\doubled$}{
            \{ $u.\token \Let v.\token$;
            $v.\doubled, u.\doubled \Let \true$; \}
        }
}
}
}
\uElse{ \tcp{$v$ or $u$ has an out-of-sync token; attempt splitting}
     \uIf{$\neg(v.\outofsync \wedge u.\outofsync)$}{
         let $v$ be the node with an out-of-sync token; \;
         \uIf{$((v.\ageinepoch < \log^{a}n) \wedge (u.\token = \notoken))$}{
            \tcp{split the $v$'s token}
             $u.\token \Let v.\token$;~ $u.\outofsync \Let \true$;\;
             $v.\ageinepoch \Let v.\ageinepoch + 1$;~  $u.\ageinepoch \Let v.\ageinepoch$;\;
             \uIf{$v.\ageinepoch = \log^{a}n$}{
                  \lFor{each $x \in \{ u,v\}$}{
                      \lIf{$x.\epochpart = 1$}{$x.\outofsync \Let \false$;}
                  }
             }
        }
   }
}
\caption{\FastExactMajorityB protocol -- a worker-worker interaction\label{FastB-worker-worker}}
\end{algorithm}

$\;$

\subsection{Chernoff bounds and Azuma-Hoeffding inequality}

In the proofs we use the following statements of Chernoff bounds
and Azuma-Hoeffding inequality (see, for example, \cite{Mitzenmacher:2005:PCR:1076315}).

\begin{theorem} {\rm [Chernoff bound]}
Let $S_n=X_1+X_2+\dots+X_n$, where $X_i$, for $i = 1, 2, \ldots, n$, are independent random variables,
$0\leq X_i\leq 1$, $\E X_i=\mu_i$ and
$\mu=\mu_1+\mu_2+\dots +\mu_m$.
Then for $0\le\varepsilon\le1$ and $\delta \ge 1$,
\begin{eqnarray}
\Pr(|S_m - \mu| \ge \varepsilon\mu)
& \leq & \exp \left\{-\frac{\varepsilon^2 \mu }{4} \right\}, \label{ChernoffA}\\
\Pr(S_m - \mu \ge \delta\mu) & \leq & \exp \left\{-\frac{\delta\mu}{3} \right\}.
\label{ChernoffB}
\end{eqnarray}
\end{theorem}

\begin{corollary}
Let $S_n=X_1+X_2+\dots+X_n$, where for $i = 1, 2, \ldots, n$,
$X_i$ is random variable such that
$X_i \in \{0, 1\}$ and $\E( X_i | X_1, X_2, \ldots, X_{i-1} ) \ge \mu_i$, and let
$\mu=\mu_1+\mu_2+\dots +\mu_m$.
Then for $0\le\varepsilon\le1$,
\begin{eqnarray}
\Pr(S_m  \le (1 - \varepsilon)\mu )
& \leq & \exp \left\{-\frac{\varepsilon^2 \mu }{4} \right\}. \label{ChernoffC}
\end{eqnarray}
\end{corollary}

\begin{theorem} {\rm [Azuma-Hoeffding inequality]}
Let a sequence of random variables $(S_0, S_1, \ldots, S_n)$
be a martingale such that $|S_{k+1} - S_k| \le c$, for each $0 \le k < n$.
Then
\begin{eqnarray}\label{Azuma}
\Pr(|S_n - S_0| \ge \D) & \leq & 2\exp \left\{-\frac{\D^2}{2nc^2} \right\}.
\end{eqnarray}
\end{theorem}

\end{document}